# Quantum Tunneling of Magnetization in Molecular Complexes with Large Spins. Effect of the Environment.

Igor TUPITSYN [a] and Bernard BARBARA

*Laboratoire de Magnétisme, Louis Néel, CNRS-BP 166, Grenoble 38042, France*
[a]*On leave of absence from Russian Research Center "Kurtchatov Institute", Moscow 123182, Russia*

## 1. Introduction

Can quantum mechanics, which determines behavior at the atomic and sub-atomic scales, be manifest at a macroscopic scale? This question, which was posed when the foundations of quantum theory were first laid, has fascinated physicists for more than seventy years.

The phenomena of superconductivity and superfluidity in helium, are quantum manifestations on a macroscopic scale. In both cases there is a macroscopic non-dissipative current of particles. More recently, quantum manifestations at scales well above the atomic scale were observed – for example, quantum tunneling of the phase in a Josephson junction, permanent currents in small conductor rings and more recently, Bose condensates. These systems, whose sizes vary from 10 to $10^5$ nm, are relatively complex; nevertheless, their properties can be described using a small number of degrees of freedom defined as a "macroscopic order parameter".

In magnetism, since the discovery of superparamagnetism by Néel, it has been known that a ferromagnetic or ferrimagnetic particle of a few nm in size can also be described with a small number of degrees of freedom, those of the magnetic moment of the single domain particle, which behaves as a small magnet (the exchange energy dominates by orienting all the moments in one direction).

The search for quantum effects at the "macroscopic scale" in magnetism started in the early seventies after it was shown that single crystals of rare-earth intermetallics ($Dy_3Al_2$, $SmCo_{3.5}Cu_{1.5}$) exhibit fast magnetic relaxation in the Kelvin range. This phenomenon was interpreted in terms of magnetization reversal by quantum tunneling, below a certain crossover temperature. The magnetization reversal of the bulk crystals being the sum of the elementary reversals of single domain blocks (nucleation of the so-called Barkhausen jumps), this type of study is equivalent to the study of single nanoparticles, but with many complications (size, energy barrier and switching field distributions, effects of domain walls, various dissipation effects…).

More recently there have been developments in various disciplines that have led to great progress in the obtention of different types of nanoparticles. In material science, magnetic materials have been produced as isolated aggregates, as deposits of aggregates, as carbon nanotubes and nanocages filled with magnetic material, as electrodeposits of magnetic material in nanoporous polycarbonate membranes, as well as dispersals in polymers. Molecular chemistry has produced molecules with giant spins and colloidal chemistry has used micelles as microreactors to make all sorts of new magnetic nanoparticles. Naturally occurring biological systems have given us ferritin and biochemistry has provided us with their artificial analogues.

Among all these systems, an exciting type of material has emerged for the study of macroscopic tunneling in magnetism - namely, molecular crystals with identical magnetic molecules. Despite their relatively large size and the dipolar interactions between their magnetic moments, these molecules clearly exhibit quantum tunneling of magnetization. The focus of this article is on two of these materials (i) the spin cluster system so-called "$Mn_{12}$-ac", which is a Manganese acetate. This system was the first to exhibit what is referred to as "resonant tunneling of magnetization". (ii) Another analogous system is the so-called "Fe$_8$", in which the same phenomenon was observed afterwards, but at lower temperature, which allows easier experimental studies.

## 2. Mn$_{12}$-acetate.

### 2.1. Experimental results.

Over the past years a lot of experimental and theoretical works were performed on molecules of [Mn$_{12}$O$_{12}$(CH$_3$COO)$_{16}$(H$_2$O)$_4$]. This molecule has a tetragonal symmetry [1] and contains a cluster of twelve Mn ions divided into two shells (four Mn$^{4+}$ ions from inner shell with spin S=3/2, surrounded by eight Mn$^{3+}$ ions from the outer shell with spin S=2) with strong antiferromagnetic couplings (frustrated triangles, see Figure 1). They form a collective ground state spin S=10 with magnetic moment M=gSμ$_B$ ≈20μ$_B$  (g≈2 is the Lande factor) [2]. These molecules are chemically identical and form a crystal with an average distance between Mn$_{12}$ molecules of the order of 15 Å [1]. Intermolecular exchange interactions are negligible and dipolar interactions are about 0.01÷0.02 K. This is much smaller than the anisotropy barrier U$_o$ of each molecule (which is about 61÷65 K [6(a),**7, 9,10**]).

Ac-susceptibility measurements of Sessoli et al. [2] and magnetization experiments of Paulsen et al. [6(a),7] show a superparamagnetic behavior with a relaxation time that obeys the Arrhenius law τ=τ$_o$exp(U$_o$/k$_B$T) (see, L. Néel in [3]) with τ$_o$=2x10$^{-7}$ sec at high temperatures (T>2.5 K) and a blocking temperature T$_B$ close to 3 K  (T$_B$≈k$_B$U$_o$/ln(t/τ$_o$)≈3.3 K for t≈1 h). Above the blocking temperature, this superparamagnetic behavior is characterized by a Curie-Weiss law with very small positive paramagnetic temperature Θ≈70 mK. This indicates the existence of weak dipole-dipole interactions between molecules. As the temperature decreases from ~3 K, the magnetization evolves from a relatively fast roughly exponential relaxation (about $10^3$ sec) to very slow non-exponential relaxation ( about $10^7$ sec at T~2.1 K ; a non-exponential behavior discovered later, will be discussed below) [6(a),7,9] (experiments [8] even give logarithmic relaxation below 1 K, but this not confirmed by more recent results [*](I. Chiorescu, R. Giraud, A. Canneschi, L. Jansen, and B. Barbara, submitted for publication). It was also found that the relaxation time $\tau$ (H) exhibits a deep minimum in zero field at T<2 K (see, for example, [7,9-11]) whereas above this temperatures $\tau$ (H) presents a maximum at 0.2 T (see Figure 2) observed, for the first time, by Paulsen et al. [7] (see, also, review paper of Barbara et al. [9]). This picture (strongly supported by dips observed in ac-susceptibility measurements by Novak et al.[10]) was interpreted as Quantum Tunneling of Magnetization (QTM) of the collective spin S=10 with a crossover temperature $T_c$~2 K, due to the resonant energy level crossing in two-well potential in longitudinal magnetic field [6(a),7,9-11,4].

The experiments reported in [6a, 9] were thoroughly repeated and confirmed in [7]. More detailed measurements performed later [4,11], showed steps in isothermal hysteresis loop (Friedman, et al. in [4] and Thomas, et al in [11]) when the field is increased in the direction opposite to the magnetization at $T<T_B$ (Figure 3, see, also [6(b)]). In the flat regions of the hysteresis loop the relaxation times were found much longer than the experimental time window (~600 sec in [11]) while in the steep regions magnetization relaxes much faster and the relaxation times can be of the order (or even less) of the experimental time window. The plot $\partial M_z/\partial H_z$ versus longitudinal filed $H_z$ (Figure 4) gives a series of peaks (with Lorentzian shape). The maxima of these peaks allows to define the values of magnetic field $H_n$ at which the magnetization steps occur : the fields $H_n$ such as $H_n \approx 0.44n$ T (n=0, 1, 2, ...). The relaxation measurements of Thomas et al. [11] show that the relaxation time oscillates with the magnetic field (Figure 5) and has deep minima (resonances) at the same values of field where the steps are observed in the hysteresis loop. These new experimental results clarify the problem of the maximum observed before at H=0.2 T [7,9,10]. As suggested by Barbara et al [9], it is effectively the first maximum of $\tau(H)$ curve. In addition, the blocking temperature $T_B$ also exhibits strong minima approximately at the fields $H_n$ (extracted from the temperature behavior of magnetization at different values of field in [4]).

All these results, interpreted as a strong evidence of QTM, were obtained at $T<T_B\approx 3.3$ K (in a longitudinal field). At the higher temperatures the magnetization relaxes too fast for quasi-static measurements and ac-susceptibility ($\chi(\omega)=\chi'(\omega)-i\chi''(\omega)$) measurements are necessary. The relaxation times can then be determined, either from the position of the maxima of the imaginary susceptibility ($\chi(\omega)$'' is maximum at $\omega\tau=1$) or from the relation $\tau=\chi''(\omega)/\omega(\chi'(\omega)-\chi'(\infty))$, where $\omega$ is the frequency of the ac-field. The higher temperature relaxation times also oscillates with minima at nearly the same field as in the low temperatures case, but the amplitude of oscillations decrease. At e.g. 10 K this amplitude is 25 times smaller than in the low-temperature case [12] (Figure 6). These measurements show that, together with the regular decrease of the relaxation time with the applied field (usual field dependence of the barrier), there is even at temperatures well above the quasi-static blocking temperature  some fraction molecules with tunneling channels, as in the low-temperature regime. This means that at high temperature, the mechanism of relaxation is intermediate between the quantum regime (where the ratio of the relaxation times at resonance and out of resonance are much smaller than 1) and the classical regimes (where this ratio goes to 1).

It was also experimentally found in [4,11] that the transition rate decreases rapidly with temperature. Therefore, if the QTM mechanism is relevant for all these experiments, then it can be understood only by Thermally Assisted QTM (introduced by Novak and Sessoli in [10] and Barbara et al. in [9]), where tunneling occurs from excited levels. As the temperature decreases, the higher levels (close to the top of the barrier) become less and less populated and the tunneling takes place between the lower levels with smaller probability, explaining the fact of the extremely long relaxation time observed below 1 K in [6(a),7,9].

However in the presence of increasing longitudinal or transverse magnetic field, the relaxation is faster and can be easily measured in the main bulk phase of $Mn_{12}$ (see very recent experiments [85]). As an example we show in Figure 7(a) some hysteresis loops obtained from torque experiments in fields up to 6 T. As in Figure 3, where measurements were done in lower fields and higher temperatures, the loops depend on temperature, but here this is true only above 0.8 K. Below this temperature the loops are independent of temperature, showing that tunneling takes place from the ground state $S=10$.

Similar results were obtained in a transverse magnetic field of about 4 T. As an example we give in the insert of Figure 7(b) the temperature dependence of the relaxation times measured at short time-scale. They are clearly independent of temperature below 0.8 K (this temperature is nearly the same as in longitudinal field by accident). This result shows that tunneling occurs here between the ground states $S=10$ and $S=-10$ of the two symmetrical wells. The main part of Figure 7(b) gives also an example of relaxation curves measured in the plateau. At short time-scale the relaxation follows a square root law, but at long time-scale it is exponential (we should note that, although the two third of the relaxation curves could be fitted by a $\ln(t)$, the short time-scale always shows the square-root law). Such a crossover between square root and exponential relaxation, occurs in all the experiments of Chiorescu et al. [85] in both longitudinal and transverese fields. Note that it was also observed in the low field and high temperature regime (see Thomas and Barbara in [57]).

## 2.2. Basic model.

To proceed with some theoretical interpretations and conclusions, first of all, one should establish a basic model describing the $Mn_{12}$ spin structure. Early experimental works of Sessoli et al. [2,13] have reported an observation of the S=10 collective spin ground state, and high frequency-EPR experiments of Barra et al. [5] have allowed to fit its results to a "giant spin" S=10 Hamiltonian which includes fourth-order anisotropy in the following form:

$$H_G = -DS_z^2 - K_\parallel S_z^4 + K_\perp (S_+^4 + S_-^4) - g\mu_B \mathbf{HS}, \qquad (1)$$

neglecting higher-order terms, with $D/k_B \approx 0.56$ K, $K_\parallel/k_B \approx 1.11 \times 10^{-3}$ K and $K_\perp/k_B \approx 2.9 \times 10^{-5}$ K. Recently, a new experimental technique, the "submillimeter spectroscopy", was applied to the $Mn_{12}$ magnetic clusters (see [14]). This technique can measure directly the energy levels which, in principle, give full information about the Hamiltonian of the system. The information can be used to tune the Hamiltonian Eq. (1) with respect to higher-order anisotropy terms.

In this model, described by Eq. (1), the fourth-order terms contribute to tunneling and, therefore, play a crucial role (discussion on the importance of neglected in Eq. (1) higher-order terms see in Section 3.2.). However, before using this Hamiltonian, it is necessary to understand how the model of collective spin S=10 ground state is stable with the increase of the temperature. In order to succeed in this task, one can try to calculate the energy levels of the entire problem (with all the couplings between 12 Mn ions as it is shown in Figure1). The dimension of the Hilbert space of such a problem is $10^8$ and, due to the obvious uncertainty in experimental determination of the anisotropy constants, any such attempts will be rather useless. To achieve a solution of the problem at least in the region of the temperatures below 150 K, Tupitsyn et al. [15] suggested to apply an idea of the "reduced" Hamiltonian. It was assumed that the largest coupling constant $J_1$ (see Figure 1), which is of the order of 200 K, locks two Mn ions with $S_1=2$ and $S_2=3/2$ into a spin state of $S_{12}=1/2$ up to the temperatures of the order of $J_1$ (the similar tactics - "dimerization" was exploited in [16,17], followed by suggestion in [13]). Using this assumption, one can apply an effective 8-spin Hamiltonian Eq. (2) to check the temperature stability of the S=10 ground state.

$$H = \frac{1}{2} \left[ \sum_{<\mu,\nu>}^{4} (C_1 \mathbf{S}_\mu \boldsymbol{\sigma}_\nu + C_\parallel S_\mu^{\ z} \sigma_\nu^{\ z}) + C_2 \sum_{<\alpha,\beta>}^{4} \boldsymbol{\sigma}_\alpha \boldsymbol{\sigma}_\beta \right] + g\mu_B \mathbf{H} \left( \sum_{\mu=1}^{4} \mathbf{S}_\mu + \sum_{\nu=1}^{4} \boldsymbol{\sigma}_\nu \right), \quad (2)$$

where $\mathbf{S}$ and $\boldsymbol{\sigma}$ distinguish four S=2 spins from the outer shell and four combined $\sigma=1/2$ spins (Figure 8). The Hilbert space of this Hamiltonian is $10^4$ and now it is possible to use the method of the exact diagonalization to get the structure of the energy levels. In spite of the fact that this Hamiltonian is "reduced" (or "truncated") it includes both exchange and anisotropy due to the second term with $C_\parallel$ and simulates the whole $Mn_{12}$ molecule up to the temperatures about 150 K at least. Note that Eq. (2) ignores Dzyaloshinskii-Moria (DM) interactions which could also be important in $Mn_{12}$ (see, for example, [18]) due to the fact that, in antiferromagnetic systems with strong couplings, this term can produce a single-ion anisotropy which can affect the tunneling (but this type of anisotropy can be "simulated" by the exchange anisotropy). In general, this is a time-reversal symmetry breaking term which can remove the Kramer's degeneracy of the energy levels (if any). Recent theoretical works (see [16] and references therein) suggest rather strong DM interaction in the $Mn_{12}$ clusters.

Since we do not know the constants $C_1$, $C_2$ and $C_\parallel$, we should use some experimental results to determine them. First of all, we have calculated the magnetization $M_z(T,H_z)$ and the susceptibility $\chi_\parallel(T,H_z)$ for different T and $H_z$. Comparing these calculated values with the experimental ones, we have tuned our coupling constants to get an agreement with the experiment. As can be seen from Figure 9, we found very good agreement for $C_1$=-85 K, $C_2$=55 K and $C_\parallel$=-7.5 K. After that we have used these constants to predict the magnetization and the susceptibility in a transverse field, $M_x(T,H_x)$ and $\chi_\perp(T,H_x)$. Figure 10 shows a good agreement with the experiment up to 150 K. Above this temperature our model is no longer valid since the pairs of spins S=2 and S=3/2 become unlocked. These calculations allow in particular to conclude about the temperature range of validity of the collective spin S=10 model. As can be seen from Figure 11, the multiplet S=9 becomes occupied above approximately 40 K and the "giant spin" model with S=10 becomes invalid. Another calculation made in the simple limit of a spin S = 10 allowed to fit very well the magnetization curves, but only up to 30-40 K, showing also that the giant spin S=10 cannot be valid above this temperature [18, 21] ; interestingly, the measured (and fitted) magnetization curves could not be distinguished from an hyperbolic tangent with S=10, showing the Ising-like character of the ground state S=10, which is occupied at temperatures below 10 K. This does not mean that the upper levels are not occupied (the phenomenon of thermally activated tunneling developed below, will show that), but simply that the weight of the ground state S=10, is dominant on the magnetization curves below 10 K. We can conclude that the tunneling in the S=10 multiplet at the temperatures above $30 \div 40$ K should be faster than in the S=9 multiplet (for example) since higher multiplets have broader and higher energy barriers. Experimentally this conclusion is confirmed by the observation of well-defined and equally spaced resonances at the temperatures above 30 K [18, 21] (instead of randomly spaced resonances). It is also interesting to note that neutron scattering experiments performed on $Mn_{12}$-ac reported the same temperature of about 40 K [19] for the transition from the S=10 ground state to S=9. This means that the model of "reduced" Hamiltonian works well in the region of temperatures not higher than $120 \div 150$ K. In the frameworks of almost the same model it was also shown by Zvezdin et al. in [20] that the susceptibility measured along a transverse field $\chi_\perp(T, H_x)$, exhibits a peak in a transverse magnetic field at about $7 \div 8$ T. This peak can be interpreted in terms of a "resonance" between the two states symmetrical with respect of the applied field. Experimentally this transition can easily be hidden by the effect of fourth order anisotropy terms on the magnetization curve, and in order to avoid thermal resonance, the temperature should be at least 0.2 K.

Now, after the temperature range of validity of the "giant spin" model is established, it is easy to estimate from Eq. (1) the critical values of the longitudinal magnetic field $H_n$ at which the intersection of energy levels occurs. The condition for the intersection of the two levels $S_z=m>0$ and $S_z=(n-m)<0$ [9,10]), simply reads as ([4,11]):

$$H_n \sim nD/g\mu_B. \qquad (3)$$

(For simplicity's sake, we neglect here all the other terms in Eq. (1) but one can take them into account in a more detailed analytical expression or numerically.) Note that, according to the fact that the steps in the hysteresis loop were discovered only when the field is increased in the direction opposite to the magnetization, we have changed a sign before the $H_z$ term. When the field is decreased, being parallel to the magnetization, there is no steps because the system is in its true ground state and there is no possibility to tunnel to another well until the field passes through zero. The value $D/k_B \approx 0.56$ K gives with Eq. (3) $H_n \approx 0.42n$ T, whereas the experimental value is $H_n \approx 0.44n$ T, as can be seen from Figure 5. At these values of the magnetic field $H_z$ the levels m>0 and (n-m)<0 come into resonance and the tunneling channels open.

The experimental measurements of the "effective barrier" height [21] allow us to identify the lowest energy level ($S_z=m_t$) where tunneling is fast enough to be recorded in a magnetization experiment. All the levels above this level have larger tunnel splitting and therefore tunnel much more rapidly. They effectively short-circuit the top of the barrier. This means that the "effective" height of the barrier in zero magnetic field is $E_{eff}(0)=D(S^2-m_t^2)$ while in non-zero longitudinal magnetic field we have $E_{eff}(H)=-E_{eff}(0)(1+H_z/2DS)^2$ (we neglected all other terms in Eq. (1) and assumed S>>1). In the high temperature regime (T~2.6÷3 K), the relaxation time behaves roughly exponentially and follows the Arrhenius law, with $E_{eff}(H)=T\ln(\tau(T,H)/\tau_0)$. Measuring the relative size of dips on the curve $E_{eff}(H)$, one can estimate $m_t$. As can be seen from Figure 12 [21], the height of the barrier in dips is reduced by about 10%, corresponding to tunneling from the levels $m_t \approx 3 \div 4$.

However if $H_z=0$, all the levels are in resonance, and tunneling can in principle take place simultaneously from all the thermally excited levels. The tunneling rate $\Gamma_0(m)$ between levels m and –m at zero temperature (and with no bias field between the levels) is approximately given by

$$\Gamma_0(m) \sim \Delta_m^2/G_0 \qquad (4)$$

(see, also [22]) where $\Delta_m$ is a tunneling splitting, $G_0$ is a level-broadening (we assume $G_0 > \Delta_m$). At nonzero temperature one has to include the thermal population of the excited levels, which gives for the tunneling rate

$$\tau^{-1} \sim \sum_m \Gamma_0(m)\exp(-E_m/k_BT), \qquad (5)$$

where $E_m$ is the energy of level with $S_z=m$. Without a transverse field in Eq. (5) there are only five terms which correspond to the tunneling between the levels linked by fourth-order term in Eq. (1) (i.e., only even values of m). From this "toy" model we can estimate the crossover temperature $T_c$ at which the thermal activation over the barrier is replaced by the tunneling through the bottom of the barrier. At $T < T_c$ the function $F_m = -E_m/k_B T + 2\ln(\Delta_m)$ has a maximum at $m=10$ while at $T>T_c$ the maximum is at $m=2$. Then, for the temperature of the crossover between these two regimes we can take $T_c \sim (E_2 - E_{10})/2\ln(\Delta_2/\Delta_{10})$. Using the method of the exact diagonalization, we have calculated from Eq. (1) the energy spectrum and the values of tunneling splitting $\Delta_{m,-m}$. Assuming that $G_0$ is independent of m (actually there is a weak dependence but we neglected it in this "toy" model), we get the crossover temperature $T_c \sim 1.3$ K for $H_z = 0$.

This simplified model is far from reality since it ignores correct description of the interactions with environment and involves only transitions $\Delta m = \pm 4$ while experimentally almost all the transitions with $\Delta m = \pm 1$ are observed (see Thomas et al. in [11]). Therefore, this model cannot be used to explain e.g. the field dependence of the relaxation time. However, even from this model, it is clear that the tunneling between the lowest levels at the temperatures about 2 K is already unfavorable. Therefore, at this temperature, the relaxation process should involve at least three steps: 1) thermal activation (by phonons) to excited levels (for example $S \rightarrow m_t$); 2) tunneling across the barrier ($m_t \rightarrow -m_t$); 3) transition to the true ground state with phonon emission ($-m_t \rightarrow -S$). This is Thermally Assisted QTM suggested for the first time in [9,10].

This was in zero transverse field. In the presence of a transverse field the situation is completely different. The splitting $\Delta_{m=S}$ is increasing proportionally to the power $2S=20$ of the ratio transverse to anisotropy field (see [71(a)], for example), the relaxation should speed up very rapidly as soon as the transverse field is a sizable fraction of the anisotropy field. Results of Barbara et al. [18] strongly suggested ground state tunneling in $Mn_{12}$ if the transverse field reaches 3-4 T. Chiorescu et al. [85] showed that in such a field, the relaxation is fast enough to be easily and completely measured. Furthermore this relaxation results from tunneling through the barrier between the ground states $S_z = \pm 10$ and becomes faster than the relaxation by thermal activation above the barrier at the crossover temperature $T_c \sim 0.8$ K. In this case where the transverse component of the applied field, is much larger than all the other transverse matrix elements, Eq. (1), in principle, can give quite satisfactory explanations of magnetic relaxation.

This is no longer the case in low transverse fields because then, even small transverse matrix elements may be relevant, in particular those resulting from the environment which is not taken into account in this section (see section 4). Nevertheless, using this equation one can still predict very interesting effects which can be experimentally observed at low temperatures and low transverse field. The tunneling splitting depends on the Haldane topological phase [32] originating from the quantum interference of possible paths (around the hard axis) between two potential minima (S and -S). This topological phase can be changed by an external magnetic field, causing the oscillations of the tunneling splitting [33,36,24,23,31]. These oscillations were already experimentally observed in the similar system $Fe_8$ [25] but we will discuss this system later. The easiest way to see the oscillations of the tunneling splitting analytically is to truncate the Hamiltonian in Eq. (1) to a simple low-energy 2 - level Hamiltonian. However, due to the presence of fourth-order terms, this task becomes rather complicated and to our knowledge the form of such a truncated Hamiltonian is not established yet. Moreover, due to the fact that higher-order anisotropy terms (up to 20th order with S=10) can contribute importantly to the value of the tunneling splitting (see discussion in section 3.2), any attempt to calculate the tunneling splitting precisely become rather pointless because higher-order anisotropy terms cannot be measured with present experimental techniques). Nevertheless, since we just want to show the principle things, we proceed with the simple biaxial Hamiltonian which includes an easy axis/easy plane anisotropy in the following way:

$$H=-DS_z^2+ES_x^2-g\mu_BH_xS_x \qquad (6)$$

The calculation of tunneling splitting for an isolated tunneling spin in the instanton technique began with the work of Enz and Shilling [71b]. For a spin tunneling in interaction with a background spins, results were first obtained by Prokof'ev and Stamp [28], with much more detailed work appearing later (see Tupitsyn et al. in [23,31]). For different aspects of this problem see also [27,34-36,71,28,72]. The obtained 2-level effective Hamiltonian reads simply as:

$$H_{eff}=2\Delta_0\tau_x\cos[\pi S-\Psi], \qquad (7)$$

where $\Delta_0$ is the tunneling splitting in zero external field (see, for example, [29,30,23,31]), $\tau_x$ is the Pauli matrix, and $\Psi$ is the Haldane topological phase:

$$\Psi=\pi g\mu_BH_x/2[E(E+D)]^{1/2}. \qquad (8)$$

This expression of $\Psi$ was given by Garg in [24] for the particular Hamiltonian where the quantization **z**-axis is choosen along the hard axis and the field is applied along this axis. Note that the result relative to the Haldane phase in [23] was given in the limit $|E|>>|D|$ i.e., $\Psi=\pi g\mu_BH_x/2E$. However, this limit does not affect the physics of the problem in general. For integer S, we get from Eq. (7):

$$\Delta_H=\langle\downarrow|H_{eff}|\uparrow\rangle=\Delta_0|\cos(\Psi)| \qquad (9)$$

whereas for half-integer S

$$\Delta_H=\Delta_0|\sin(\Psi)| \qquad (10)$$

Eq. (9) and (10) clearly show oscillations of the tunnel splitting as a function of the transverse magnetic field, together with the parity effect (see [35-38,23,31]). This last effect tells that half-integer spin does not tunnel in zero transverse field. Note that a magnetic field along the easy axis (or along the medium axis which is **y**-axis in this case) does not produce any oscillations and this can be seen mathematically from the fact that the effective tunneling splitting $\Delta_H$ in this case behaves like $\Delta_0|\cosh(\Psi)|$ (for details see Tupitsyn et al. [23,31]). Eq. (7) gives the general effective 2-level Hamiltonian describing tunneling with a transverse magnetic field ( in thiscase $\Psi$ $=\Psi_H+i\Psi_M$ is complex ; the contributions from hard and medium axes are denoted by $\Psi_H$ and $\Psi_M$, respectively).

In the case of Mn$_{12}$, where the lowest-order transverse anisotropy term is of fourth-order (higher order terms have not yet been determined), we can write instead of Eq. (8), (for H$_y$=0):

$$\Psi=\pi g\mu_B H_x/T_x(D,K_\parallel,K_\perp,S), \tag{11}$$

where $T_x(D,K_\parallel,K_\perp,S)$, the period of the oscillations along the **x**-axis, can be calculated numerically from Eq. (1). Note that this equation has two hard axes (**x** and **y**) which are equivalent. This means that oscillations with the same period should be seen along both directions. Using the exact diagonalization method, we have calculated the tunnel splitting $\Delta_{m,-m}$ for different m and the results can be seen in Figure 13. First of all, it is easy to understand that the tunneling splitting has a non-zero value in zero transverse magnetic field, only for levels with even values of m (which is related to the fourth-order anisotropy term). For all the other levels (with odd m) the tunneling splitting is non-zero only if the magnetic field is finite (which also produces transitions due to S$_+$ and S$_-$ ). In order to see easily why the oscillations can be seen only in a finite region of transverse magnetic field (from –H$_c$ to H$_c$), let us again forget for a moment about the fourth order term and return to Eq. (6). Combining two nondiagonal terms, we get the function $A(\theta,\varphi)=(\sin(\theta)\cos(\varphi)-H_x/2ES)^2$. When H$_x$<H$_c$=2ES, $A(\theta,\varphi)$ as a function of $\varphi$ has two local minima at nonzero $\varphi$. Since the Haldane phase is nothing else but the area on unit sphere enclosed by two possible paths between two minima $S_z=S$ and $S_z=-S$, in the case of $H_x<H_c$ the topological phase (which is an imaginary part of the instanton action) has nonzero value and the changes in $H_x$ results in oscillations of $\Delta$. However for $H_x>H_c=2ES$, the function $A(\theta,\varphi)$ has only one local minimum at $\varphi=0$, meaning that both paths joining the states $S$ and $-S$ coincide (up to the $\varphi-$fluctuations of trajectories which renormalize the value of $H_c$ to $[2E(E+D)]^{1/2}S$). The area enclosed by these two paths, i.e. the imaginary part of the instanton action is zero and, therefore, there are no more oscillations of the tunneling splitting (see also [24]).

It is important to note that the number of zeros of the $\Delta_{m,-m}(H_\perp)$ function strongly depends on the symmetry of the anisotropy terms. For the Hamiltonian of Eq. (1) (case of $Mn_{12}$) where the lowest transverse anisotropy term is of fourth order, the number of zeros for even values of $m$ (along positive or negative direction of field) should not exceed the number of times ($v=1,\ldots,5$) the operator $S.^4$ has to be applied to the state $S_z=m$ to reach the state $S_z=-m$. (However, this is only valid for $K_\perp>0$ ; if $K_\perp<0$, the axes **x** and **y** are no longer the hard axes and $\Delta(H_\perp)$ obviously would not show the oscillations along these directions.) The tunneling splitting for odd $m$ has an additional zero at zero magnetic field. In this case the chain of operators $S.^4$ or $S_+^4$ which should be applied on the state $|m\rangle$ to reach the state $|-m\rangle$ must be completed by additional operators S. (or $S_+$) which come from magnetic field term (i.e. $|-m\rangle = (S.)^2(S.^4)^v|m\rangle$ with $v=0,\ldots,4$). The same situation should occur for the tunneling between the levels involved into the resonance by applying non zero longitudinal magnetic field (say, levels $S_z=-m$ and $S_z=m-n$). Only the levels linked by $(S_+^4)^v$ or $(S.^4)^v$ ($v=0,\ldots,4$) can have non zero tunneling splitting in zero transverse field (see Figure 14). Moreover, it is easy to see from this figure how the period of the first oscillation decreases when $n$ increases inside the group of curves $(-10, 9) \div (-10, 7)$ (or $(-10, 5) \div (-10, 3)$). To reach the states $S_z$=9, 8, 7 from the state $S_z$=-10 we should apply $S_-^4$ four times and then, to complete the chain linking the mentioned states, apply $S_-$ three, two or one time.

In addition, if we apply the transverse magnetic field at different azimuth angles $\varphi$ and increase $\varphi$ from 0, the amplitude of the oscillations becomes smaller and vanishes at $\varphi=\pi/4$ since in the case of fourth-order anisotropy $\pi/4$-axis is an easy axis. However, increasing $\varphi$ from $\pi/4$ to $\pi/2$, one can see the same curves. This is quite obvious from the symmetry of the problem (see Figure 15). This is the case when $K_\perp>0$. If $K_\perp<0$, as we noted already, **x** and **y** axes are no longer the hard ones. Two axes along the directions of $\varphi=\pm\pi/4$ become harder for the system. Some oscillations along these directions should be observed.

## 3. Fe₈ octanuclear iron (III) complexes.

### 3.1. Experimental results.

Another molecule, so-called Fe$_8$, with the chemical formula [Fe$_8$O$_2$(OH)$_{12}$(tacn)$_6$]$^{8+}$, where tacn represents the organic ligand triazacyclononane, is under intensive investigations at the present time. It contains eight iron (III) ions (S=5/2) with strong antiferromagnetic couplings between ions [39] (see Figure 16). Similarly to Mn$_{12}$ they form an uncompensated S=10 collective ground state. Four Fe$_8$ ions in the middle of the molecule are in the so-called "butterfly arrangement". This system is nearly orthorombic with a strong Ising-like anisotropy and giving an energy barrier of about 24 K (i.e., about one-third of that in Mn$_{12}$) [39,40]. An analysis of the magnetic susceptibility v.s. temperature shows that only the levels S>8 are populated near 10 K [39]. (As for Mn$_{12}$ (see above), one must say that upper levels are also occupied, but the weight of the ground state S=10, dominates the susceptibility.)

Magnetic relaxation experiments have been performed in Fe$_8$, following the same procedure shown above, for Mn$_{12}$. The relaxation rate becomes temperature-independent below 0.35 K, as was shown by Sangregorio et al. in [41]. This can also be seen in the hysteresis loop recently shown for a single crystal [42] (Figure 17). Like in Mn$_{12}$, equally-spaced steps were observed, but with a smaller spacing : $\Delta$H$\approx$0.22 T instead of 0.44 T. In [41] it was found that at these values of field (H$_n$$\approx$0.22n T), the relaxation becomes much faster than that in the plateaus. These observations give a second example of tunneling across the anisotropy energy barrier, when the levels from the opposite sides of the barrier come into resonance. At low temperature and without transverse field, the ratio of the relaxation time measured at resonance and out of resonance is larger than in Mn$_{12}$ by one or two decades, showing that the relaxation in Fe$_8$ is, in this case, faster than in Mn$_{12}$ by the same factor (one or two decades).



(However, as mentioned above, in the presence of a transverse field of a few Tesla the relaxation of $Mn_{12}$ becomes much faster ; if the transverse field is of 3-4 T, ground state tunneling between S=+10 and S=-10 is observed below the crossover temperature $T_c \approx 0.8$ K [85].) In $Fe_8$, ground state tunneling occurs below $T_c \approx 0.35$ K, but in this case it is not necessary to apply a transverse field. To fit the relaxation data, a stretched exponential law $M(t)=M(0)exp[-(t/\tau)^{\beta(T)}]$ (with $\beta(T)$ increasing from approximately $0.4 \div 0.5$ (below 0.4 K) to nearly 1 at T~1 K) was used [41]. This law was also observed for measurements on an oriented crystal by Ohm et al. (see [43,56]). At temperatures below 0.35 K and at short times, the best fit of the data is a square root behavior $M(t)=M(0)[1-(t/\tau_{short})]^{1/2}$ (see Figure 18). This law was, in fact, predicted by the theory of Prokof'ev and Stamp [44] for the relaxation due to tunneling at the bottom of the barrier, at short times and low temperatures, with initial magnetization near saturation. Later these measurements were repeated at lower temperatures (T=40 mK) by Wernsdorfer et al. and the square root law was confirmed [45]. Note, however, that this law was also found in a zero-field cooled annealed sample, which is then allowed to relax in a finite field, i.e. near zero magnetization. A similar effect is also observed in $Mn_{12}$ [85] (see below). This needs further theoretical investigations (see [91]).

In conclusion, in $Fe_8$ the relaxation goes in the pure quantum regime via the tunneling through the barrier between the states $S_z=\pm 10$ at temperatures $T \leq T_c \approx 0.35$ K and zero transverse field. Similarly, in $Mn_{12}$ the ground state tunneling was also observed, however a transverse field of 3-4 T must be applied to achieve this result ; the crossover temperature is then larger than in $Fe_8$ ($T_c \approx 0.8$ K for $Mn_{12}$). The application of such a transverse field in $Fe_8$ would makes the relaxation so fast that it would be impossible to measure it (unless by EPR). At higher temperatures ac-susceptibility experiments of Caneschi et al. [42] show peaks similar to those observed in $Mn_{12}$



[12,21], but they are more pronounced than in $Mn_{12}$, from which we may conclude that $m_t$ (m-value for the barrier short-cut) is larger in $Fe_8$. However one should keep in mind that the collective spin S=10 of this molecule breaks down near 10 K, suggesting that the peaks observed at high temperature (7 K) come, for a large part, from multi-spins tunneling. This is corroborated by the fact that the peaks are not regularly separated and that their mean separation looks closer to 0.14 T than to the 0.22 T of the spin S=10.

### 3.2. Basic model.

High-Frequency EPR technique was applied by Barra et al. [40] to investigate the magnetic anisotropy in $Fe_8$ molecules. They found a biaxial anisotropy described by the Hamiltonian $H=-DS_z^2+E(S_x^2-S_y^2)$. More recent neutron spectroscopy experiments (see Caciuffo et al. in [46]) have reported the presence of fourth-order term. According to these experimental data, the Hamiltonian for the $Fe_8$ molecule can be written as follows:

$$H_G = -D_0S_z^2 + E_0(S_x^2-S_y^2) + K_\perp(S_+^4 + S_-^4) - g\mu_B\mathbf{HS} \qquad (11)$$

which is equivalent to:

$$H_G = -DS_z^2 + ES_x^2 + K_\perp(S_+^4 + S_-^4) - g\mu_B\mathbf{HS} \qquad (12)$$

where $D/k_B=(D_0-E_0)/k_B \approx 0.23$ K , $E/k_B=2E_0/k_B \approx 0.092$ K and $K_\perp/k_B \approx -2.9\text{x}10^{-5}$ K ($g \approx 2$, see [40], [46]). Eq. (11) and (12) are valid only in the temperature range where the collective spin S=10 can be defined, i.e. at temperatures lower than 10 K [39] (this temperature is 40 K in $Mn_{12}$).

Similarly to the case of the $Mn_{12}$ molecule we can estimate the value of the longitudinal magnetic fields at which the levels from the opposite sides of the barrier (say, $S_z=m$ and $S_z=n-m$) come into resonance. Eq. (12) gives:

$$H_n=(nD/g\mu_B)[1+E/2D]. \qquad (13)$$



(we again neglected the fourth-order term). The contribution E/2D compensates the difference between D and $D_0$ (compare with Eq. (3)) and therefore one must have $D_0=D[1+E/2D]$, which is true and identical to $D=D_0-E_0$ (see above). Eq. (13) gives $H_n \approx 0.205n$ (in Tesla), whereas the experimental value is 0.22n T. The above considerations show that we may have some uncertainty in both the actual values of the constants and in the types of the anisotropy terms included into the Hamiltonians. As was pointed out in [44,47], higher-order transverse anisotropy terms (even with very small constants) can make an important contribution to the value of the tunneling splitting. This can be seen easily from the perturbation theory (for the lowest order perturbation approach for the tunneling splitting see [48,49,71]). Following [71] a simple form for the tunnel splitting can be written [21] (we omit here dependence on value of S): $\Delta_{-m,m} \sim D(K_{\perp p}/D)^{2m/P}$ (where p is the order of the anisotropy term in the Hamiltonian). As an example, Eq. (12) gives for m=S=10 and p=2 or p=4, $\Delta_{-10,10} \sim D(E/D)^{10}$ or $\Delta_{10,-10} \sim D(K_\perp/D)^5$. The contribution of e.g. the 10th order term gives already $D(K_{\perp 10}/D)^2$. Unless the quasi-exponential increase of $\Delta$ with p, a divergence of higher orders of the tunneling splitting is forbidden due to the fast decrease of the constants $K_{\perp p}$ (these constants are directly connected to the crystal field parameters which we know, decrease very rapidly with the expansion order p). Nevertheless, $\Delta$ depends on the value of $K_{\perp p}$ in a so crucial way that all the terms up to 20th order can be important. However, the parameters $K_{\perp p}$ of the higher orders (except p=4) are experimentally unmeasurable at the present time. This makes very problematic any quantitative calculations of the tunneling splitting from Eq. (12). The actual behavior of $\Delta_{m,n-m}(H_x)$ strongly depends on the values of the anisotropy constant. Any uncertainties in these constants cause the changes in the period and amplitude of the oscillations in the transverse magnetic field (see Figures 13-15). Furthermore, number of the



oscillations (which are confined in a given field interval) itself depends on the relative values of the constants. In the case of the simplest example of the biaxial anisotropy of Eq. (6) the field interval for oscillations is [-$H_c$,$H_c$], where $H_c \sim [2E(E+D)]^{1/2}$. Consider the tunnel state where $|-10\rangle$ and $|10\rangle$ are admixed, with splitting $\Delta_{10,-10}$. In order to reach the state $|-10\rangle$ starting from the state $|10\rangle$, the operator $S_+^P$ must be applied 20/p times (for even p). The main consequence, is that the number of oscillations depend directly on the symmetry of the anisotropy: $\nu(p=2)=10$ oscillations for a second order anisotropy or $\nu(p=4)=5$ oscillations for a fourth order one. It is clear that if $K_\perp$ increases from zero, the transition from $\nu(p=2)=10$ to $\nu(p=4)=5$ will not be discontinuous, because both periodicity will be involved in the interference. If $K_\perp$ is negligible (in Eq. 12), only the second order term contributes to the splitting. However, as $K_\perp$ increases above some critical value $K_{\perp c} \sim 2E/S^2$ (with $K_\perp > 0$) its contribution becomes dominant and $\nu(p=2)=10$ decreases progressively to the value $\nu(p=2)=4$ (as in the case of Eq. (1)). Simultaneously, the period and the amplitude of oscillations increase. One must note that in all that the changes in anisotropy constants are not small: the transition between $\nu(p=2)$ and $\nu(p=4)$ occurs when the parameters of consecutive orders in Eq. (12), are nearly the same. In the above example, $K_{\perp c} \sim 2E/S^2 \sim 1.8 \ 10^{-3}$ K is about $10^2$ times the real value of $K_\perp$! Unless very unusual values of the crystal field parameters, such a transition could not be observed. In any case, this situation is not stable since the splitting resulting from such large $K_\perp$ is itself very large: $\Delta_c \sim 7 \ 10^{-3}$ K, i.e. nine orders of magnitude larger than the actual splitting (given above; let us note casually the huge effect of $K_\perp$ on $\Delta$). Comparing the value $\Delta_c$ to the splitting given e.g. Figure 22, it is not difficult to imagine that the energy spectrum will changes dramatically and all the levels with definite $S_z$ will be completely admixed ($S_z$ will not be conserved). In the opposite



case ($K_\perp < 0$), the increase of $|K_\perp|$ above $K_{\perp c}$ leads to a disappearing of the oscillations because the **x**-axis is no longer the hard axis of the system (as well for the **y**-axis, due to the tetragonal symmetry of Eq. (12)).

Any other anisotropy terms of order higher than four, could in principle, change the number, the period and the amplitude of the oscillations. These oscillations have been recently observed in the system $Fe_8$ by Wernsdorfer et al. [25]. Figure 19 shows the measured tunneling splitting as a function of transverse magnetic field at different azimuth angles $\varphi$. Using the Hamiltonian (Eq. (12)), we have calculated the tunnel splitting in a transverse field by simple diagonalisation of the 21x21 matrix. We get similar oscillations as in [25] with the following parameters: $D/k_B = 0.23$ K , $E/k_B = 0.094$ K and $K_\perp/k_B = -3.28 \times 10^{-5}$ K. There are some differences between the curves calculated for $\varphi = 0°$ and the measured one (the calculated curve is sharper near the nodes, and the experimental curve shows some increase of the value of $\Delta_{-10,10}(H_x)$ in the nodes). However, the curve calculated for $\varphi = 1°$ (shown in Figure 20), is more similar to the experimental one. This suggests some misorientations in the experiments. The reason for this suggestion is connected with the fact that an increase of the magnetic field along the medium axis (which is the **y**-axis in our case) produces an increase of the tunneling splitting which obeys the relation $\Delta_0 |\cosh(\Phi_y)|$ (where $\Phi_y$ is the Haldane phase for the magnetic field $H_y$ and $\Delta_0$ is the tunneling splitting in zero field). The simplest way to simulate this non-zero component of $H_y$ field (not unique however) is to introduce some "misalignment" angle between the **x**-axis and the direction of the applied magnetic field. One cause could be the mosaic which is always present in molecular crystals of this type, which is also of the order of 1°. However this gives random misorientations between say +0.5° and –0.5°. Such a mosaic should modify the curve $\Delta_{-10,10}(H_x)$, but in a distributed way, and this could contribute to the observed broadening of the nodes.



Another origin for misorientation comes from the triclinic symmetry of the Fe$_8$ molecule; in this case symmetry lowering should induce new crystal field parameters and new contributions to the tunneling splitting, which are in first approximation taken into account by field H$_y$. In what follows, we will use the value of this "misalignment" angle $\theta_m = 1°$.

The calculated $\Delta_{-10,10}(H_x)$ with zero longitudinal field ( but $\theta_m = 1°$), can be seen in Figure 20. The period of the oscillations is about 0.41 T in agreement with the experimental value. The curves with the larger value of $\phi$ (up to $\pi/2$) clearly show that the oscillations disappear (in agreement with the experimental behavior) when the direction of the applied field approaches the medium axis (**y**). As for the absolute value of $\Delta_{-10,10}(H_x)$ and the shape of the last oscillation, the agreement is not very good. These discrepancies may give an indication that the Hamiltonian Eq. (12) is not quite satisfactory with respect to the unknown higher anisotropy terms. It would always be possible to choose the value of these terms up to highest order (20) to get the best fit. One might also consider the effects of couplings to the environment. In order to give an idea of the influence of high order terms, we show Figure 21, the effect of the fourth-order term added to the second order one. Starting from Eq. (12), we have calculated numerically the period of oscillations T$_H$, for different values of K$_\perp$. When this term is null i.e. when the period is given by Eq. (8), the value of T$_H$ is about one half the measured one. The value of the fourth order term K$_\perp$ =-2.9 10$^{-5}$ K, allows to recover the measured period.  One could also take another value of K$_\perp$, determined independently, and fit the period on the eight order term (for example). This is a kind of unproductive activity (at the present time), and we stop here this discussion.

At higher values of K$_\perp$ the period depends on K$_\perp$ almost linearly but the value of interest ($|K_\perp| \sim 3 \div 4 \times 10^{-5}$ K ) is in the nonlinear region. In the absence of longitudinal



field, ($\Delta_{-m,m-n}$ with m=10 and n=0), it is easy to interpolate $T_H$ by some simple formula even in this region, but all the other anisotropy constants also have uncertainty in their values. To tabulate $T_H$ as a function of all available anisotropy constants, we have used the combination of the method of exact diagonalization with the method of polynomial interpolations. The final formulas are valid in quite a wide region of the values of the anisotropy constants (in Kelvins):

$$D/k_B \in [-0.06,-0.45]; \; E/k_B \in [0.05,0.13]; \; K_\perp/k_B \in [-0.8\text{x}10^{-5},-5.2\text{x}10^{-5}] \qquad (14)$$

and can be written as follows:

$$T_H=(2k_B/g\mu_B)[E(E+D)]^{1/2} \sum_{\mu,\nu=1}^{3} X_\mu{}^K \, G_{\mu,\nu} \, X_\nu{}^E, \qquad (15)$$

where

$$G_{\mu,\nu}= \sum_{\alpha,\beta=1}^{3} X_\alpha{}^\nu \, \Omega^\mu{}_{\alpha,\beta} \, X_\beta{}^D; \qquad (16)$$

$$\mathbf{X}^1=\begin{pmatrix} 1 \\ 0 \\ 0 \end{pmatrix}; \; \mathbf{X}^2=\begin{pmatrix} 0 \\ 1 \\ 0 \end{pmatrix}; \; \mathbf{X}^3=\begin{pmatrix} 0 \\ 0 \\ 1 \end{pmatrix}; \; \mathbf{X}^D=\begin{pmatrix} D^2 S^2 \\ DS \\ 1 \end{pmatrix}; \; \mathbf{X}^E=\begin{pmatrix} E^2 S^2 \\ ES \\ 1 \end{pmatrix}; \; \mathbf{X}^K=\begin{pmatrix} K_\perp{}^2 S^8 \\ K_\perp S^4 \\ 1 \end{pmatrix}; \qquad (17)$$

$$\Omega^1=\begin{pmatrix} -0.2027 & -0.7838 & -1.6242 \\ 0.3877 & 1.3462 & 4.0468 \\ -0.1251 & -0.0644 & -2.6684 \end{pmatrix};$$

$$\Omega^2=\begin{pmatrix} -0.1790 & -0.5550 & -1.6797 \\ 0.3449 & 0.8759 & 4.4967 \\ -0.1031 & 0.2321 & -3.6051 \end{pmatrix}; \qquad (18)$$

$$\Omega^3=\begin{pmatrix} -0.0498 & -0.2483 & -0.0030 \\ 0.1155 & 0.6170 & 0.0108 \\ -0.0653 & -0.3996 & 0.9928 \end{pmatrix}.$$

For the general case (n>0), one has to come back to full numerical calculations. The period depends on the values of n and m (Figure 22 and 23 a,b). Figure 22 shows



the tunneling splitting $\Delta_{m,-m}$ (in Kelvins) as function of $H_x$ (in Tesla) in zero longitudinal magnetic field for the azimuth angle $\varphi=0$ (with "misalignment" angle $\theta_m=1^{\circ}$). Due to the presence of the second-order term in Eq. (12) all the $\Delta_{m,-m}$ have nonzero values in zero transverse field. As we discussed above, negative $K_\perp$ tends to decrease the number of the oscillations along the hard **x**-axis because it makes **x**-direction easier for quasiclassical motion of the giant spin. This effect leads to the decrease of the imaginary part of the instanton action (by decreasing the area on the unit sphere, enclosed by two possible paths joining quasiclassical minima – S and S). In this case it is difficult to calculate the exact number of oscillation. First of all, an analytical solution of the problem in the presence of the fourth-order term is not easy, in particular the perturbation theory cannot be applied in large transverse magnetic field. The instanton technique and WKB approximations give non-analytical solutions which require numerical calculations (there is no reason to apply it since we already have an answer given by the method of exact diagonalization). As mentioned above, the only conclusion that we can make here is that the number of oscillations is determined by the exponents $\alpha,\beta,\gamma$ in the chain of operators $(S_\pm^1)^\alpha(S_\pm^2)^\beta(S_\pm^4)^\gamma$ which should be applied to the state $|m\rangle$ to reach the state $|n\text{-}m\rangle$. The operator $S_\pm$ comes from the transverse magnetic field term, $S_\pm^2$ and $S_\pm^4$ come from the second and fourth order anisotropy terms, respectively. It is clear that the whole picture is defined by the combined symmetries of the anisotropy terms. This is apparent in Figure 22, with the change of the period with m and the shape of the last oscillation. Note in particular that the quantities $\Delta_{10,-10}$, $\Delta_{8,-8}$ and $\Delta_{6,-6}$ (with the difference $\delta S_z=4$ in the lengths of the chains connecting $|m\rangle$ and $|-m\rangle$ states) show a similar behavior. The same situation takes place for $\Delta_{9,-9}$, $\Delta_{7,-7}$ and $\Delta_{5,-5}$. For smaller m, the structure of interference is also affected



by the admixing of states with different m ($S_z$ is not a good quantum number for the Hamiltonian Eq. (12)).

The same conclusions can be drawn, in the presence of a longitudinal field. The functions $\Delta_{-10,10-n}(H_x)$ calculated for $\varphi=0$, are plotted in Figure 23 (with n even – (a) or odd – (b)). The resonant longitudinal field for particular n is defined by Eq. (13) (of course, it was necessary to tune the value of the field around these $H_n$). The levels $S_z=-10$ and $S_z=10-n$ with odd values of n cannot be linked in zero transverse field (there are no matrix elements between them) and therefore they have a zero tunneling splitting at $H_x=0$. The levels $S_z=-10$ and $S_z=10-n$ with even values of n can be linked even in zero transverse field, and they have a finite tunnel splitting. (As an example for n=1 and n=2, one has to apply $S_{\pm}$ and $S_{\pm}^2$ respectively.) The curves calculated in Figure 23, can be compared with the ones measured in $Fe_8$ [25] and plotted in Figure 24. The absolute value of the tunneling splitting, again, is not in the best possible agreement with the experiment. Unfortunately, the authors of [25] did not show behavior of the curves with n=1,2 (and with larger n) at higher values of transverse magnetic fields. The comparison between measured and calculated curves in the region of high fields can help to determine which types of anisotropy are important in the "giant spin" Hamiltonian. In other words, measurements of the tunnel splitting as a function of a magnetic field can be used to determine the crystal field parameters of the "giant spin" Hamiltonian. This method should be quite sensitive to the anisotropy of the higher orders

In addition, we would like to stress that the actual behavior of the tunneling splitting in the region of the nodes is very sensitive to the environment. This point will be discussed below in the part of this review related to the low temperature limit of the relaxation of the magnetization (Environmental effects).



















This figure "fig20.gif" is available in "gif" format from:



This figure "fig21.gif" is available in "gif" format from:













This figure "fig32.gif" is available in "gif" format from:






**REFERENCES.**

[1] T. Lis, *Acta Crystallogr. B 36*, 2042 (1980).

[2] R. Sessoli, D. Gatteschi, A. Caneschi, and M.A. Novak, *Nature 365*, 141 (1993).

[3] L. Néel, *Ann. Geophys. 5*, 99 (1949).

[4] J.R. Friedman, M.P. Sarachik, J. Tejada, and R. Ziolo, *Phys. Rev. Lett. 76*, 20 (1996).

[5] A.L. Barra, D. Gatteschi, and R. Sessoli, *Phys. Rev. B 56*, 8192 (1996).

[6] (a) C. Paulsen, J.G. Park, B. Barbara, R. Sessoli, and A. Caneschi, *J. Magn. Magn. Mater. 140-144*, 379 (1995); (b) C. Paulsen, J.G. Park, B. Barbara, R. Sessoli, and A. Caneschi, *J. Magn. Magn. Mater. 140-144*, 1891 (1995)

[7] C. Paulsen and J.P. Park, in L. Gunter, B. Barbara (Eds.), *Quantum tunneling of Magnetization – QTM'94*, NATO ASI Series E: Applied Science, Vol. **301**, Kluwer, Dordrecht, 189 (1995).

[8] J.A.A.J. Perenboom, J.S. Brooks, S. Hill, T. Hathaway, and N.S. Dalal, *Phys. Rev. B 58*, 333 (1998).

[9] B. Barbara, W. Wernsdorfer, L.C. Sampaio, J.G. Park, C. Puulsen, M.A. Novak, R. Ferre, D. Mailly, R. Sessoli, A. Caneschi, K. Hasselbach, A. Benoit, and L. Thomas, *J. Magn. Magn. Mater. 140-144*, 1825 (1995).

[10] M.A. Novak, R. Sessoli, in L. Gunter, B. Barbara (Eds.), *Quantum tunneling of Magnetization – QTM'94*, NATO ASI Series E: Applied Science, Vol. **301**, Kluwer, Dordrecht, 171 (1995).

[11] L. Thomas, F. Lionti, R. Ballou, D. Gatteschi, R. Sessoli, and B. Barbara, *Nature 383*, 145 (1996).

[12] F. Lionti, L. Thomas, R. Ballou, B. Barbara, R. Sessoli, and D. Gatteschi *J. Appl. Phys. 81 (8)*, 4608 (1997).





[13] R. Sessoli, H.L. Tsai, A.R. Shake, S. Wang, J.B. Vincent, K. Folting, D. Gatteschi, G. Christou, and D.N. Hendrickson, *J. Am. Chem. Soc.* **115**, 1804 (1993).

[14] A.A. Mukhin, V.D. Travkin, A.K. Zvezdin, S.P. Lebedev, A. Caneschi and D. Gatteschi, *Europhys. Lett.* **44** *(6),* 778 (1998).

[15] I. Tupitsyn , P.C.E. Stamp, B. Barbara, L. Thomas, (preprint 1996).

[16] M.I. Katsnelson, V.V. Dobrovitski, B.N. Harmon, cond-mat/9807176.

[17] A.K. Zvezdin, A.I. Popov, Sov. *Phys. JETP* **82**, 1140 (1996).

[18] B. Barbara, L. Thomas, F. Lionti, I. Chiorescu, A. Sulpice, *J. Mang. Magn. Mater.* ***177-181***, 1324 (1998).

[19] M. Hennion, L. Pardi, I. Mirebeau, E. Suard, R. Sessoli, and A. Caneschi, *Phys. Rev. B* **56**, 8819 (1997).

[20] A.K. Zvezdin, V.V. Dobrovitski, B.N. Harmon, and M.I. Katsnelson, *Phys. Rev. B* **58**, R14723 (1998).

[21] B. Barbara, L. Thomas, F. Lionti, I. Chiorescu, A. Sulpice, *J. Magn. Magn. Mater.* (accepted for publication).

[22] L.D. Landau and E.M. Lifshits, *"Quantum mechanics"*, 1965 (Oxford, Pergamon).

[23] I.S. Tupitsyn, N.V. Prokof'ev, and P.C.E. Stamp, *Int. J. Mod. Phys. B* **11**, 2901 (1997).

[24] A. Garg, *Europhys. Lett.* **22**, 205 (1993); cond-mat/9906203.

[25] W. Wernsdorfer and R. Sessoli, *Science* **284**, 133 (1999).

[26] D.A. Garanin, E.M. Chudnovsky, *Phys. Rev. B* **56**, 111002 (1997).

[27] I.Y. Korenblit, E.F. Shender, *Sov. Phys. JETP* **48**, 937 (1978).

[28] N.V. Prokof'ev and P.C.E. Stamp*, J. Phys. Cond. Mat.* **5**, L663 (1993).

[29] A.I.Vainshetejn, V.I. Zakharov, V.N. Novikov, and M.A. Shifman. *Usp. Fiz. Nauk*, **136**, 553 (1982) [*Sov. Phys. Usp.* **25**, 195 (1982).





[30] S.N. Burmistrov and L.B. Dubovski, *Preprint IAE-3881/1 [in Russian]*, Kurchatov Institute of Atomic Energy, Moscow (1984).

[31] I. Tupitsyn, *JETP Lett., 67*, 28 (1998). (cond-mat/9712302).

[32] F.D.M. Haldane, *Phys. Rev. Lett. 50*, 1153 (1983).

[33] E.N. Borachek and I.V. Krive, *Phys. Rev. B 46*, 14559 (1992).

[34] A.S. Ioselevich, *JETP Lett. 45*, 69, 3232 (1987).

[35] J. von Delf and C. Henley, *Phys. Rev. Lett. 69*, 3236 (1992).

[36] D. Loss, D.P. di Vinchenzo, and G. Grinstein, *Phys. Rev. Lett., 69*, 3232 (1992).

[37] H.B. Braun and D. Loss, *Phys. Rev. B 53*, 3237 (1996).

[38] A. Garg, *Phys.Rev. Lett. 74*, 1458 (1995).

[39] C. Delfs, D. Gatteschi, L. Pardi, R. Sessoli, K. Wieghardt, and D. Hanke, *Inorg. Chem. 32,* 3099 (1993).

[40] A.L. Barra, P. Debrunner, D. Gatteshi, Ch. E. Shultz, and R. Sessoli, *Europhys Lett., 35 (2),* 133 (1996).

[41] C. Sangregorio, T. Ohm, C. Paulsen, R. Sessoli, and D. Gatteschi, *Phys. Rev. Lett., 78*, 4645 (1997).

[42] A. Caneschi, D. Gatteschi, C. Sangregorio, R. Sessoli, L. Sorace, A., Cornia, M.N. Novak, C. Paulsen, and W. Wernsdorfer, *J. Mang. Magn. Mater. 200*, (accepted for publication, 1999).

[43] T. Ohm, C. Sangregorio, and C. Paulsen, *Eur. Phys. J. B 6*, 195 (1998).

[44] N.V. Prokof'ev and P.C.E. Stamp, *Phys. Rev. Lett., 80*, 5794 (1998).

[45] W. Wernsdorfer, T. Ohm, C. Sangregorio, R. Sessoli, D. Mailly, and C. Paulsen, *Phys. Rev. Lett. 82*, 3903 (1999).

[46] R. Caciuffo, G. Amoretti, and A. Murani, *Phys. Rev. Lett., 81*, 4744 (1998).

[47] N.V. Prokof'ev and P.C.E. Stamp, *J. Low. Temp. Phys., 104*, 143 (1996).





[48] D.A. Garanin, *J. Phys. A: Math. Gen.* **24**, L61 (1991).

[49] F. Hartmann-Boutron, *J. Phys. I France* **5**, 1281 (1995).

[50] F. Luis, J. Bartolome, and J. Fernandez, *Phys. Rev. B* **57**, 505 (1998).

[51] A. Fort, A. Rettori, J. Villain, D. Gatteschi, and R. Sessoli, *Phys. Rev. Lett., 80*, 612 (1998).

[52] M.N. Leuenberger and D. Loss, *Europhys Lett., 46 (5)*, 692 (1999); cond-mat/9907154.

[53] W. Wernsdorfer, R. Sessoli, and D. Gatteschi, *Europhys. Lett., 47 (2)*, 254 (1999).

[54] D. V. Berkov, *Phys. Rev. B* **53**, 731, (1996).

[55] P.W. Anderson, *Phys. Rev.* **82**, 342 (1951).

[56] T. Ohm, C. Sangregorio, and C. Paulsen, *J. Low. Temp. Phys., 113*, 1141 (1998).

[57] L. Thomas and B. Barbara, *Phys. Rev. Lett*., **83**, 2398 (1999).

[58] L. Thomas and B. Barbara, *J. Low. Temp. Phys.* **113**, 1055 (1998).

[59] L. Thomas, Ph.D. *Thesis*, Université Joseph Fourier, Grenoble, France, 1997.

[60] J. R. Friedman, M.P. Sarachik, R. Ziolo, Phys. Rev. B 58, R14729 (1998). Note that in their determination of the line-width, the authors have mentioned that the procedure of the preparation of the sample (powder of the crystallites oriented in a field) is equivalent to that used in [4]. However, according to [4] (see Figure 1 of this paper), the magnetization at zero field is less than 0.25 of its saturated value. It is quite clear that the crystallites in this powder are not so well oriented. In fact this result is very strange because this fraction of 0.25 is even smaller than the 0.5 value which is expected for a completely disordered ensemble of uniaxial spins. This will of course, and at least, increases the experimental errors on the line-width.

[61] P.C.E. Stamp, in S. Tomsovic (Editor), *'Tunneling in Complex Systems",* Word Scientific Publishing, Singapore, 101 (1998).





[62] J.M. Hernandez, X.X. Zhang, F. Luis, J. Tejada, J.R. Friedman, M.P. Sarachak, and R. Ziolo, *Phys. Rev. B 55*, 5858 (1997).

[63] J. Villain, F. Hartmann-Bourtron, R. Sessoli, and A. Rettori, *Europhys. Lett., 27*, 159 (1994).

[64] P. Politi, A. Rettori, F. Hartmann-Bourtron, and J. Villain, *Phys. Rev. Lett. 75*, 537 (1995).

[65] F. Hartmann-Bourtron, P. Politi, and J. Villain, *Int. J. Mod. Phys. 10*, 2577 (1996).

[66] D.A. Garanin and E. M. Chudnovsky, *Phys. Rev. B 56*, 11102 (1997).

[67] A. L. Burin, N.V. Prokof'ev, and P.C. E. Stamp, *Phys. Rev. Lett. 76*, 3040 (1996).

[68] V.V. Dobrovitski and A.K. Zvezdin, *Europhys. Lett., 38 (5)*, 377 (1997).

[69] L. Gunter, *Europhys. Lett., 39*, 1 (1997).

[70] A. Caneschi, T. Ohm, C. Paulsen, D. Rovai, C. Sangregorio, and R. Sessoli, *J. Magn. Magn. Mater. 177-181,* 1330 (1998).

[71] (a) J.L. van Hemmen and S. Suto, in L. Gunter, B. Barbara (Eds.), *Quantum tunneling of Magnetization – QTM'94*, NATO ASI Series E: Applied Science, Vol. **301**, Kluwer, Dordrecht, 189 (1995). *Europhys. Lett., 1*, 481 (1986); *Physics B 141*, 37 (1986); (b) M. Enz and R . Shilling, *J. Phys. C 19*, L711 (1986); (c) E.M. Chudnovsky and L. Gunther. *Phys. Rev. Lett. 60*, 661 (1988); (d) P.C.E. Stamp, E.M. Chudnovsky, and B. Barbara, *Int. J. Mod. Phys. B 6*, 1355 (1992).

[72] N.V. Prokof'ev and P.C.E. Stamp, *cond-mat/9511011*.

[73] N.V. Prokof'ev and P.C.E. Stamp, in L. Gunter, B. Barbara (Eds.), *Quantum tunneling of Magnetization – QTM'94*, NATO ASI Series E: Applied Science, Vol. **301**, Kluwer, Dordrecht, 347 (1995).

[74] Yu. Kagan, L.A. Maksimov, Sov. *Phys. JETP 52 (4)*, 688 (1980).

[75] L.D. Landau and E.M. Lifshits, *"Statistical physics"*, 1965 (Oxford, Pergamon).





[76] A.M. Gomes, M.N. Novak, R. Sessoli, A. Caneschi, and D. Gatteschi, *Phys. Rev. B 57*, 5021 (1998).

[77] N.V. Prokofev and P.C.E. Stamp, *J. Low. Temp. Phys. 113*, 1147 (1998).

[78] J.F. Fernandez, F. Luis, and J. Bartolome, *Phys. Rev. Lett., 80*, 5659 (1998).

[79] W. Wernsdorfer, unpublished.

[80] W. Wernsdorfer, T. Ohm, C. Sangregorio, R. Sessoli, D. Gatteschi, and C. Paulsen, (to appear in *Physica B*).

[81] A. Cuccoli, A. Fort, A. Rettori, E. Adam, and J. Villain, *cond-mat/9905273*.

[82] C. Zener, Proc. R. Soc. *London A 137*, 696 (1932); S. Miyashita, *J. Phys. Soc. Jpn. 64*, 3207 (1995).

[83] W. Wernsdorfer, R. Sessoli, A. Caneshi, D. Gatteshi, and A. Cornia, (to appear in *J. Appl. Phys.*)

[84] G. Rose, P.C.E. Stamp, and I.S. Tupitsyn, (submitted to *Phys. Rev. Lett.*, 1999).

[85] I. Chiorescu, R. Giraud, A. Canneschi, L. Jansen, and B. Barbara, (to be published)

[86] A.Lascialfari, Z. H. Jang, F. Borsa, P. Carretta, and D. Gatteschi, *Phys. Rev. Lett. 82*, 3773, (1998)

[87] T. Goto, T. Kubo, T. Koshiba, U. Fujii, A. Oyamada, J. Arai, T, Takeda, K. Awaga. (submitted to *Physica B*, **LT22**. 18 Sept. 1999), and private comm.

[88] Zhang and Widom, *JMMM 122*, 119 (1993)

[89] R. Ferré and B. Barbara, (1995) not published

[90] A. Wurger, J. Phys.: Condens. Matter 10 10075, (1998)

[91] This case can be handled by a generalization of the original zero-field theory [44]. N.V. Prokof'ev and P.C.E. Stamp (to be published).


**FIGURE CAPTIONS.**



**Figure 1**. Interaction scheme of Mn$_{12}$ molecule.

**Figure 2**. Variation of $\tau(H_z)$ measured in [7,9] at T>T$_c$. Note that, in this figure (which we take from [9]), the value of Tln($\tau(H_z)/\tau_0$) is plotted to show also that U$_0$(H$_z$) (effective barrier height) deviates from ~DS$^2$(1+H$_z$/2DS)$^2$.

**Figure 3**. Hysteresis loops of Mn$_{12}$ with the field along z-axis from [11].

**Figure 4**. Field variation of the derivative $\partial M_z/\partial H_z$ taken at 1.9 K along the hysteresis loop of the single monocrystal [12]. The sharp peaks correspond to the magnetization jumps and flat regions correspond to the plateau located between the jumps. The continuous line is a fit to the Lorentzian peaks centered at H$_n$.

**Figure 5**. Relaxation times at temperature 1.9 K versus H$_z$ obtained in [11] from repeated measurements for given H$_z$ and T on the hysteresis loop. The insert shows the relaxation time drops against inverse temperature.

**Figure 6**. Field dependence of the relaxation time $\tau=\chi''(\omega)/\omega(\chi'(\omega)-\chi'(\infty))$ from AC susceptibility measurements [12]. The dashed line represents the fit of the thermal activation background and the continuous curve is a fit of thermally activated resonance dips.

**Figure 7**. (a): Hysteresis loops obtained on a single crystal of Mn$_{12}$ (main phase) from torque experiments [85] performed at temperatures between 1.3 and 0.4 K. The magnetization was first saturated in a large positive field. The field was then decreased to zero and reversed. Data points were taken between -2.5 and -5.5 T. The sweeping field velocity was equal to 10.8 mT/s. The steps amplitude depend on temperature, but only above 0.5 - 0.8 K. Below this temperature the hysteresis loops are independent of temperature, suggesting tunneling from the ground-state S=10. (b): an example of magnetic relaxation experiments performed near the maximum of the first resonance in the presence of a transverse field of about 4 T, on a single crystal of Mn$_{12}$. The data



were taken after saturation in a positive field and fast application of the field to the top of the resonance. This curve shows that the relaxation follows a square root law at short time-scale and an exponential law at long timescale. Insert: temperature dependence of the square root relaxation time. Below 0.8 K, the relaxation time is independent of temperature showing the existence of grounde-state tunneling between S=10 and S=-10 in this bulk phase of $Mn_{12}$.

**Figure 8**. Simplified coupling scheme of $Mn_{12}$ molecule.

**Figure 9**. (a) Temperature dependence of longitudinal susceptibility $\chi_\parallel$. Triangles are experimental results from [59]. Dashed line is $\chi_\parallel T$ calculated from Eq. (2) with the contribution only from the energy levels up to 100 K starting from the ground state (85 levels). Solid line is $\chi_\parallel T$ calculated from Eq. (2) with the contribution from all the energy levels ($10^4$ levels). (b) Dots are magnetization curves $<M_z>/M_s$ versus longitudinal field measured in [59] at different temperatures. Solid curves are the same curves calculated from Eq. (2).

**Figure 10**. (a) Temperature dependence of transverse susceptibility $\chi_\perp$. Triangles are experimental results from [59]. Dotted line is $\chi_\perp T$ calculated from Eq. (2) with the contribution only from the energy levels up to 100 K. Dashed line is the same but with the contribution from energy levels up to 500 K (2982 levels). Solid line is $\chi_\perp T$ calculated from Eq. (2) with the contribution from the energy levels up to 1000 K (8362 levels). (b) Dots are magnetization curves $<M_x>/M_s$ versus transverse field measured in [59] at different temperatures. Solid curves are the same curves calculated from Eq. (2).

**Figure 11**. Energy spectrum, calculated from Eq. (2) up to 180 K. The stars show a parabolic behavior ($A(S^2-S_z^2)$ where $-S \leq S_z \leq S$ and A=0.627 K) of S=10 multiplet.



**Figure 12**. Effective energy barrier of $Mn_{12}$ molecule versus longitudinal field from [21]. The solid, dashed and dotted curves represent the law $E_{eff}(H)=\Delta_{eff}=-\Delta(1+H_z/2DS)^2$ for different values of the barrier height $U_0=\Delta$.

**Figure 13**. Tunneling splitting $\Delta_{-m,m}$ ($Mn_{12}$) calculated from Eq. (1) for different m versus transverse magnetic field.

**Figure 14**. Tunneling splitting $\Delta_{-10,10-n}$ ($Mn_{12}$) calculated from Eq. (1) for different n versus transverse magnetic field.

**Figure 15**. Tunneling splitting $\Delta_{10,-10}$ ($Mn_{12}$) calculated from Eq. (1) for the different values of the azimuth angle φ versus transverse magnetic field.

**Figure 16**. Interaction scheme of $Fe_8$ molecule.

**Figure 17**. Hysteresis loops recorded on a single crystal of $Fe_8$ molecules in [42] at different temperatures and at constant sweep rate $\partial H/\partial t=0.14$ T/sec.

**Figure 18**. Square root of time relaxation curves for $Fe_8$ single crystal measured in [43] (see also [56]) at 150 mK for $M_{in}=M_s$. The insert shows the distribution of $\tau_{sqrt}^{-1}$ extracted from the above data as a function of field.

**Figure 19**. Ground state tunneling splitting $\Delta_{10,-10}$ measured for several azimuth angles φ versus transverse magnetic field [25].

**Figure 20**. Ground state tunneling splitting $\Delta_{10,-10}$ calculated from Eq. (12) for several azimuth angles φ versus transverse magnetic field.

**Figure 21**. The period of the oscillations of the ground state tunneling splitting $\Delta_{10,-10}$ calculated from Eq. (12) as a function of the fourth-order anisotropy constant $K_\perp$.

**Figure 22**. Tunneling splitting $\Delta_{-m,m}$ ($Fe_8$) calculated from Eq. (12) for different m versus transverse magnetic field.



**Figure 23**. Tunneling splitting $\Delta_{-10,10-n}$ (Fe$_8$) calculated from Eq. (12) for different n versus transverse magnetic field. (a) – even and (b) – odd values of n.

**Figure 24**. Tunneling splitting $\Delta_{-10,10-n}$ (Fe$_8$) measured for n=0,1,2 [25].

**Figure 25**. Quantum hole digging in the initial distribution of internal fields [53]. Note that these measurements were performed on a minor species of Mn$_{12}$. (For details about the different species of Mn$_{12}$, see [53,42] and references therein.)

**Figure 26.** The digging time dependence of the hole shape in Mn$_{12}$ from [53].

**Figure 27**. Temperature dependence of the hyperfine line-width ($\sigma_{hyp}$) in the crystal of Mn$_{12}$ [79].

**Figure 28**. Tunneling distribution in Fe$_8$ (which, according to [44], is proportional to the distribution of P($\xi_H$) of the internal bias field $\xi_H$) for annealed sample measured in [45]. The insert enlarges the region of the fields around the hole.

**Figure 29**. Dependence of the hole shape (in initial distribution of the dipolar fields in Fe$_8$) on degree of annealing [45]. At initial magnetization $M_{in} < |0.5M_s|$ the hole becomes independent on future annealing and gives the line-width $\sigma_{hyp} \sim 1.2 \div 1.6$ mT.

**Figure 30**. Temperature dependence of the hyperfine line-width width ($\sigma_{hyp}$) in the crystal of Fe$_8$ [80].

**Figure 31.** Time decay of the magnetization in Mn$_{12}$, plotted as a function of the square root of the time [58]. Lines show linear fits.

**Figure 32**. Square root of time relaxation curves for Fe$_8$ crystal measured at 40 mK in [45] for $M_{in}$=0.

**Figure 33**. The value of the tunneling splitting $\Delta_{S,n-S}$ versus transverse magnetic field $H_x$ ($\varphi$=0°) around the first node at the different values of the initial magnetization $M_{in}$ [83].

**Figure 34**. The value of **γ** calculated from the Eqs. (12) and (61) in the case of strongly annealed sample (the Gaussian distribution of the dipolar fields). (a) - **γ** versus



transverse magnetic field at different values of Gaussian half-width $E_D$; (b) – enlarged region of transverse magnetic field around the first node; (c) – the value of $\boldsymbol{\gamma}$ in the node versus $E_D^2$.

**Figure 35.** The value of $\boldsymbol{\gamma}$ versus transverse magnetic field calculated with small "misalignment" angle $\theta_m = 1°$ at different values of $E_D$.



**4**. **Environmental effects.**

At the present time, current theories are not able to explain relaxation experiments in all the temperature ranges. Therefore, it is reasonable to consider the low temperatures (ground-state tunneling regime) and the higher temperatures (thermally assisted tunneling regime) separately. The first is better understood by the theory of Prokof'ev and Stamp [44,47,72,73,77] while progresses in understanding of the second are more related to the works of Luis et al. [50], Fort et al. [51] and Leuenberger et al. [52]. Since the barrier height in $Fe_8$ is about three times smaller than in $Mn_{12}$, the values of the tunneling splitting are larger for all states. In particular one expects a barrier cur-off at larger $m_t$ in the high temperature regime, and faster relaxation in the low temperature regime. These circumstances make $Fe_8$ molecules very attractive for the study of the ground-state tunneling regime (i.e. pure quantum regime). The thermally assisted regime is usually (historically) the main subject of investigations on $Mn_{12}$ molecules.

## 4.1. Experimental picture.

To deal with relaxation one has to take into account the environmental effects. The environment of a molecule is essentially constituted of ensembles of bosons and fermions coupled to the spin of the molecule. In real molecular systems the most important environmental effects come from phonons, nuclear spins, dipolar fields. The environment is able to absorb finite variations of energy and angular momentum. This is extremely important because the non-conservation of these quantities can forbid the tunneling. The environment is also responsible for the broadening of resonance lines. The shape of these lines is usually obtained from the plot of $dM_z/dH_z$ v.s. $H_z$, as this was first defined in Ref. [11]; in $Mn_{12}$ nearly Lorentzian resonance line-shapes were



found of width (40mT to 100 mT). Furthermore, as mentioned by Barbara et al. [18,11], tunneling resonance could not have been detected with the poor field resolution of conventional SQUID magnetometers, if resonance lines were not importantly broadened and since pure phonons broadening is too small by a factor of 100 [65], broadening of magnetic origin must play the major role (the line-with in the absence of environment is ~ tunnel splitting $\Delta$) . This is true at low temperature only because the effect of phonons is obviously dominant in the thermally assisted regime. Very similar transitions broadening were observed in $Fe_8$ by Ohm et al.; at low temperature they are of the order of $12 \div 15$ mT [43].

Now we describe the origin of the resonance line-width in molecular crystals, in the two limits of high and low temperatures (see also [21] for qualitative description from experiments). In the thermally activated regime (where $\Delta$ at the bottom of barrier can reach values of the order of 0.5 K) the line-width is at least equal to the value of the tunneling splitting. The resonance line in this regime must be homogeneously broadened with Lorentzian shape due to equilibrated spin-phonons transitions, and this was observed in $Mn_{12}$ (see e.g. [11, 12, 60]). Note that, deviations from a Lorentzian line-shape can be obtained depending on the magnetic history of the sample, and this is because the width of distributions of dipolar and hyperfine fields are of the same order with the tunneling splitting. It might be surprising to see that magnetic history, can be important in a basic effect such as tunneling. In fact the magnetic history is always important when irreversible process come to play a role, and this is the case here. It is well known that frozen distributions of hyperfine and dipolar fields (quenched from the super-paramagnetic state) have Gaussian distributions and if the temperature is low enough to prevent fast spin reorganizations. In this case resonance lines will be inhomogeneously broadened, with no Lorentzian and eventually Gaussian line-shapes.



In the low temperature regime, where thermal fluctuations are essentially frozen, tunneling will only be possible through internal fields fluctuations, of amplitude $H_f$. These fluctuations are due to nuclear spins in weak dipole-dipole interactions, and with strong hyperfine interactions with electronic spins. Internal field fluctuations involve spin-spin quantum dynamics of the considered system which can be tested on either electronic or nuclear spins. In $Mn_{12}$ each ion has a nuclear spin, whereas in $Fe_8$ this is the case for only 2% of the ions ($Fe^{57}$). However, each $Fe_8$ molecule contains 120 hydrogen, 18 nitrogen and 8 bromine atoms, which all have nonzero nuclear magnetic moment. They produce (together with the dipolar moment of each molecule) the fluctuating internal field acting on each molecule. The amplitude $H_f$ of these fluctuating fields are extremely small (they can be evaluated from NMR experiments [86]; e.g. protons in $Fe_8$ give $H_f \sim 1.4$ mT [21]). Quantum tunneling being only possible within this range of fluctuating fields, spin reversals from S to –S will occur only in a narrow energy window of width $H_f$, "digging a hole" in the initial distribution of internal fields. The frozen distribution of internal fields (resulting from electronic and nuclear spins which are not affected by the quantum dynamics) could be obtained from the measured $dM_z/dH_z$ v.s. $H_z$, in sweeping the longitudinal field, as we said above. However, contrary to the high temperatures case, the distribution which is probed here for each value of $H_z$, is inhomogeneous, and it will be possible to probe this distribution as long as $H_f$ will remain much smaller than its width. This will be the case unless the temperature increases to the point where $H_f$ becomes of the order of the total local field distribution. The cross-over between inhomogeneous Gaussian-like to homogeneous Lorentzian-like distributions is when the temperature becomes large enough to equilibrate the spin system. In $Mn_{12}$, the tunneling window $H_f$ was evaluated assuming oscillations of the mean dipolar and hyperfine field with $\Delta m = \Delta I = \pm 1$ (see [21,87]).



Recently low temperature experiments (0.04÷0.3 K) of Wernsdorfer et al. [53], using the "hole digging" technique [45] based on the theory of Prokof'ev and Stamp [44], allowed to measure the tunneling window. In this theory the relaxation rate of the magnetization $\Gamma_{sqrt}(H)$ is proportional to the distribution of internal bias field $P(\xi_H)$. On the experiment, during the digging time $t_{dig}$ a small fraction of molecules (that are in resonance in applied external digging field $H_{dig}$) tunnel reversing the direction of their magnetization. This causes rapid transitions of molecules which are close to the resonance around $H_{dig}$. Such transitions are effectively "digging a hole" in an initial distribution of internal fields (see Figure 25). The hole widens in time, depending on the digging time $t_{dig}$. The hole-width (which is obtained by the linear interpolation to $t_{dig}=0$) gives an intrinsic broadening of the nuclear fields $\sigma_{hyp}$ (see Figure 26). It was found [53] that the width of the Gaussian distribution of hyperfine fields $\sigma_{hyp} \approx 12$ mT which is larger than the value of the tunneling splitting $\Delta_{10,-10}$ by orders of magnitude. $\sigma_{hyp}$ is the temperature independent up to approximately 0.4 K and then starts to increase (see Figure 27). (The last fact needs further theoretical investigations since the square-root theory [44] does not work at so high temperatures.) We should note, however, that these "hole digging" experiments were performed on the minority phase of $M_{12}$ and in the presence of a magnetic field this system can have a lower crossover temperature. Low temperature studies of main phase of $M_{12}$ were recently possible. A large magnetic field was applied to get ride of the minority phase (see [85]).

Similarly to $M_{12}$ it was observed in $Fe_8$ that the first maximum in the relaxation rate is not in zero applied field, but at 8 mT. As in $M_{12}$ this was attributed to the effect of internal fields [41]. This was in a powder sample. In a single crystal it was found that the resonance width (of about 12÷15 mT) is as in $M_{12}$ and for the same reason (intermolecular dipolar interaction), orders of magnitude larger than expected without



environment or with phonons only [18,11,65]. A difference between these two systems is that the first resonance is observed in a negative field in $M_{12}$ and in a positive one in $Fe_8$. This was interpreted as a consequence of the competition between the demagnetizing field -NM (shape-dependent, where N is the demagnetizing factor) and the local Lorentz field $+(4\pi/3)M$ [21]. Since the $M_{12}$ crystals are elongated, the demagnetizing field (which is antiparallel to M) is smaller than the Lorentz field and the internal field is parallel to the magnetization M: one has to apply a negative field to cancel the internal field. The situation is just opposite with $Fe_8$. This discussion was relative to the most probable values of internal fields. Now, we know that internal fields are distributed with a width, measured in both systems (see above). Hole-digging experiments similar to those described above for impurity phase of $M_{12}$ were performed in $Fe_8$ [45], also at low temperature. It was found that, for thermally annealed sample (down to values of magnetization of about $-0.2M_s$, where $M_s$ is the saturated magnetization), the distribution $P(\xi_H)$ of the internal bias field $\xi_H$ is very accurately described by a Gaussian function (Figure 28). As a matter of fact such a distribution is expected from theoretical models, but only in the limit of high spin concentrations (dipolar field distribution of dense set of randomly oriented spins, see [88,54]). In the other limit of dilute static dipoles, the distribution has to be Lorentzian, as this was shown by Anderson in [55]. A study for different concentrations was done showing that in between these two limits more complicated and sometimes bi-modal distributions could occur [89]. Interestingly, in Figure 28 the maximum of $P(\xi_H)$ is shifted from H=0. The shift is even larger than in [41] and this is because the experiment was done at lower temperature: the magnetization and thus the local field are larger (the sample was not absolutely annealed) The width of this distribution ($\sigma_{dip}$) was found to be of the order of 50 mT).



The relaxation of the molecules, like in $Mn_{12}$, digs a hole in $P(\xi_H)$ at the value of the applied field. The width and depth of the hole change with the waiting time i.e. with the time during which the field is applied (see insert in Figure 28). For initial magnetizations close to the saturation the hole is large and asymmetric, whereas for the initial magnetization less than $|0.5M_s|$ it becomes symmetric, independent on further annealing (on initial magnetization) and has a width of about $1.2 \div 1.6$ mT (see Figure 29). Similarly to $Mn_{12}$ this line-width is temperature independent up to 0.4 K and then it starts to increase (see Figure 30). Some values of the hole width were predicted by Prokof'ev and Stamp at zero Kelvin [44]: 0.3 mT for $Fe_8$ and 25 mT for $Mn_{12}$, which are close to the measured ones, before these experiments were performed and come (in their theory) from the nuclear spins. It was shown that the dipolar fields produce the bias $\xi_H$ which a few orders of magnitude larger than the value of the tunneling splitting and, therefore, can block the tunneling. However, the fast nuclear dynamics (transverse relaxation in nuclear subsystem or, in other words, $T_2$ processes) broadens the resonance line and opens a channel for tunneling. This explains the origin of the hole (its width is defined by the width of the distribution of the hyperfine fields) in the field-dependent relaxation rate (which is proportional to the total distribution of internal fields).

All these low-T experiments performed on $Fe_8$ [41,43,45,53,56]] demonstrate the square-root relaxation law at short times (about first 100 sec, see e.g. Fig. 17). According to the theory of Prokof'ev and Stamp, this law comes from the time-dependent distribution of fluctuating internal dipolar fields in a sample. Fast initial transitions change the total distribution of internal fields across the sample that can push some molecules out of resonance but bring other molecules into the resonance everywhere in the sample and allow continuous relaxation. For longer times the



experimental relaxation data are better fitted by the stretched exponential law with $\beta \sim 0.4$ [41,43].

In $Mn_{12}$, Thomas and Barbara [57-59,21] show that the relaxation behaves non-exponentially even at relatively high temperatures (up to 2.8 K) that can be understood as a consequence of intermolecular dipolar coupling. (At low temperatures non-exponential relaxation was seen in many experiments. See, for example, [4,6,7,11,53,62]). The exponential relaxation can take place only in the limit of non-interacting (with each-other and with nuclear thermostat) molecules when we can simply write $dM(t)/dt = -\tau^{-1}M(t)$. In the case of interacting molecules, the right-hand side of this equation contains some different value $M^*(t) = f(M(t))$ which gives deviation from the exponential law (note that a simple distribution of relaxation times gives also a non exponential behavior). As it was found in [57], below approximately 1.7 K the relaxation follows a square-root law with characteristic time $\tau$ weakly dependent on temperature whereas at the temperatures above 2.5 K the characteristic time follows the Arrhenius law, but the relaxation is still non-exponential because of the dipolar interaction between molecules. To fit the experimental relaxation curves, it was necessary to use the stretched exponential law with $\beta(T) < 1$. Below 1.9 K $\beta(T)$ is approximately constant near 0.5. However, at low temperatures the stretched exponential was not really satisfactory. As can be seen from Figure 31, the square-root law fits all the experiments up to $\sim 1.7 \div 1.8$ K (see also Thomas and Barbara in [58]).

At higher temperatures, between 2 K and 2.8 K, dipolar interactions play a less fundamental role, but their influence is quite observable, e.g. from the fact that the relaxation is clearly non exponential (the exponent of a stretch exponential fit increases from 0.5 to 1, see [57-59,21]) and also by the shift of the maximum of the relaxation curve in magnetic field. This shift, found by Thomas and Barbara in $Mn_{12}$ at different



temperatures [57-59,21], is due to the evolution of the internal field with temperature. They suggested [57,21] that this shift can produce a square root like relaxation laws of origin different from the Prokof'ev and Stamp law. Instead of being at zero Kelvin, this law, with temperature-dependent relaxation times, can hold at high temperature, even if the system is equilibrated. This suggests that a quantitative description of the Thermally Assisted Regime in $Mn_{12}$ requires to take into account interactions with the spins environment (not only with the spin-phonon interaction). This also explains [see 21] why the resonance lines have a Lorentzian shape [11,12] although the relaxation is not exponential. Note that even if the observed resonances can be well fitted to Lorentzian law [11,12] (and not to the Gaussian one associated with hyperfine and dipolar fields [60]) this does not mean that hyperfine and dipolar interactions are not relevant. The widths of the dipolar fields and hyperfine fields distribution in $Mn_{12}$ are comparable, and these fields can participate in the resonance broadening but their non-Lorentzian character is hidden by the long Lorentzian tails (and experimental error bars). The fact that the observed Lorentzian line-shape does not prove that the system of molecules is equilibrated, was show in [21] where it is shown that the shape of the resonance depends on the history of the sample (field cooled or zero-field cooled, quenched or not quenched). Finaly one should mention that the published line-widths depend on the authors; they are going from 20 mT [60] to 30 mT [53] or 35mT [12]. In [21] a continuous increase of the resonances line-width was observed with the index of the resonance at temperatures close to 3 K. This effect being not symmetrical with the magnetization state M=0, one cannot infer the dipolar field distribution here. It may rather be due to the intrinsic line-width $\Delta$, which is at the higher temperatures (when the tunneling takes place at the top of the barrier) of the order of 0.5 K (which is larger than the other contributions [21]) and which increases in average with the applied field. This



observation confirms that, even if dipolar interactions and hyperfine interactions (with longitudinal $T_1$ relaxation processes, driven by the fast dipolar processes [44,47,61]) are essential in the range 2 K to 2.8 K, at the highest temperatures (say between 2.8 and 3 K) the intrinsic line-width $\Delta$ plays the most important role. In this case i.e. at temperatures rather close to the blocking temperature (3.0 K), the transitions are homogeneously broadened and the observed Lorentzian line-shapes really attest for thermal recovery of the spin-phonons system. (Note, that fast dipolar flip-flop processes are also able to provide thermal quasi-equilibrium on each side of the barrier, producing, simultaneously, rapidly fluctuating fields acting on each molecule.) In this limit the spin-phonon interaction becomes predominant leading to the Thermally Assisted Tunneling Regime.

The observation of history-dependent peak shapes in $Mn_{12}$ at relatively low temperature (1.6 K) [21], leads us to ask the question of the validity of experimental determinations of Lorentzian line-shapes. To extract this distribution from the measured relaxation curve vs field, it is necessary to assume an analytical law to fit the data. As an example, Friedman et al. in [60] fitted the experimental curve to $\exp(-t/\tau)$. However, as this is shown above [57-59, 21,53] the relaxation deviates clearly from an exponential even at these high temperatures (2-2.6 K). This must lead to large errors in the relaxation time (obtained at each field) and, therefore, in the resonance line-width. Note that even in exponential regime, error bars on the equilibrium magnetization $M_{eq}$ in $M(t)=M_{eq}+(M_{in}-M_{eq})\exp(-t/\tau)$ lead to uncertainties in the line-shape. The value of $M_{eq}$ can be obtained accurately only in the case of very fast relaxation, i.e. either very close to the blocking temperature of 3 K (and in this case the observation of a Lorentzian form makes no doubt) or in the presence of a large transverse magnetic field which increases the tunneling gap [85].



In conclusion, each molecule can tunnel under the effect of fast fluctuating field originating from dipolar and hyperfine interactions, and also of spins-phonons interactions at higher temperatures. It is only in this last case (where the transition width is intrinsic) the spins and phonons is equilibrated (no hole in the spin energy distribution).



## 4.2. Thermally Assisted Tunneling Regime.

The first attempt to apply phonons-based mechanisms was made by Villain et al. in [63] and Politi et al. in [64] before the resonant tunneling in $Mn_{12}$ was confirmed experimentally. Ref. [64] considered the giant spin model for single $Mn_{12}$ molecule in longitudinal magnetic field (Eq. (1) with no $S_z^4$ term) with coupling to the acoustic phonons in the form:

$$H_{sp-ph} = \sum_q \ (\eta/2N_uM_u\omega_q)^{-1/2}[iqV_q(\mathbf{S})C_q^+ - iqV_q^+(\mathbf{S})C_q], \qquad (19)$$

where $V_q(\mathbf{S}) \sim D(S_xS_z + S_zS_x)$, $N_u$ is the number of unit cells, $M_u$ is the mass per unit cell and $C_q$ is the phonon annihilation operator. They have ignored the possibility of the tunneling (due to the forth-order anisotropy term) at the top of the barrier (as well as at the bottom) and have got the result $\tau^{-1} \sim (SH_z)^3$ for the relaxation rate, in contradiction with the experiments which demonstrated the existence of a minimum of the relaxation time near $H_z=0$ instead of maximum (see Figure 5). Later, this theory was extended by including the interactions with nuclear spins (the possibility of tunneling was ignored again and only the longitudinal part of hyperfine interaction was considered) [65]. It was found that combination of these two mechanisms (strictly speaking, the sum of two different curves) can give a minimum of the relaxation time at zero field. This theory was also unable to describe the resonant behavior of the relaxation curve. Phonon-mediated tunneling relaxation was considered in the theory of Garanin and Chudnovsky (see [66]) which involved also random hyperfine fields. However they ignored higher-order anisotropy terms in the Hamiltonian and based their calculations on the perturbation theory for smaall transverse field $H_x$. All these models are qualitatively or quantitatively in contradiction with the experiments. Nevertheless, they are not incorrect and they show (directly or indirectly) the importance of the hyperfine interaction as well as of the intermolecular dipolar coupling (see, for example, Burin et al. in [67]). The



problem is that to get a quantitatively correct answer, it is necessary to consider (even at temperatures around 2 K) that the tunneling effect involves all the interactions of the spin of each molecule with the environment of spins and phonons. This is very difficult task. For example, Dobrovitski and Zvezdin in [68] have concluded that a correct description of the jumps width of the hysteresis loop (see Figure 3) requires to take into account fluctuating internal fields since they found for pure giant spin Hamiltonian a huge discrepancy with the experimental results. However, they just limited themselves to the suggestion that the origin of this field could be of dipole-dipole nature. They estimated an average value for jumps width using a Gaussian distribution of fluctuating field. In some sense similar calculations were made by Gunther in [69] who calculated the width of the jump in the hysteresis loop and concluded that it is necessary to involve the dynamical transverse magnetic fields to avoid the discrepancies with the experiment.

The first theory taking into account the fourth order anisotropy terms is of Luis et al. [50]. They presented a theory of resonant quantum tunneling of large spins through thermally activated states which includes: phonon-mediated transitions between the states m and m' with m-m'= $\delta$m=$\pm$ 1 (in a simplest form $S_xS_z+S_zS_x$), resonant tunneling due to fourth-order anisotropy terms, and transverse magnetic field. They assumed that the transverse magnetic field originates from the combined action of dipolar and hyperfine fields. They concluded that, of course, internal fields can not explain the minima on the relaxation curve alone but, together with fourth-order terms, these fields can account for the experimentally observed behavior of the relaxation in $Mn_{12}$. To obtain the life-time of the excited levels (due to phonon-mediated transitions) they applied a standard master equation. Finally, the spin-relaxation rate was averaged over a Gaussian distribution of longitudinal dipolar fields (together with hyperfine ones). Despite the fact of the inequivalence in treatment of tunneling between the resonance



states and phonon-mediated transitions, this model was the first one which described (qualitatively) the hysteresis loop of $Mn_{12}$ [11], as well as other experimental results at high temperatures (T>2.5 K). It is important to note, however, that in this model the magnetization relaxes always exponentially "after a brief non-exponential relaxation" (see also the same authors in [78]). However, as we have seen above, the magnetization relaxes non-exponentially, unless the temperature is very close to the blocking temperature (about 3 K).

Recently Fort et al. in [51] have improved calculations of Villain et al. [63-65] by adding tunneling through the top of the barrier. They derived the master equation:

$$dN_m/dt = \sum_{p=1}^{2} N_{m-p}\gamma_{m-p}{}^m + \sum_{p=1}^{2} N_{m+p}\gamma_{m+p}{}^m - N_m \sum_{p=1}^{2} (\gamma_m{}^{m-p} + \gamma_m{}^{m+p}) - (N_{-m}-N_m)\Gamma_m, \quad (20)$$

where $N_m$ is the number of molecules in spin state $|m\rangle$, $\gamma_m{}^p$ – the spin-phonon relaxation rate from state $|m\rangle$ to state $|p\rangle$ and $\Gamma_m$ is the tunneling relaxation rate from state $|m\rangle$ to state $|-m\rangle$. This equation includes the phonon mediated transitions with $\delta m = \pm 1, \pm 2$ together with tunneling between the states $|m\rangle$ and $|-m\rangle$. Using Eq. (20) authors concentrated on the investigations of the first resonance ($H_z=0$) on relaxation curve (see Figure 5). They noted that in the region of the validity of their theory ($H_z<3$ kOe) the relaxation rate calculated at the temperatures T=2.8 K and T=2.97 K are in good agreement with the experimental one and this agreement between the theory and the experiment becomes better at higher temperature. They also noted that a general treatment of this problem requires to incorporate other interactions (dipolar, random fields, etc.) since the model with fourth-order anisotropy term gives only transitions with $\delta m = \pm 4$. To modify the theory, it was suggested to include some transverse fields (of any nature) which obey selection rule $\delta m = \pm 1$ (i.e., contain terms like $S_x$ or $S_y$).



Very recently Leuenberger and Loss [52] presented a theory of the relaxation in $Mn_{12}$ at high temperatures (2 K and higher) based on thermally assisted spin tunneling in weak transverse magnetic field. They solved the standard master equation (for the reduced density matrix $\rho(t)$) which includes both the resonance tunneling due to fourth-order anisotropy term (as well as transverse fields) and phonon-induced transitions with $\delta m = \pm 1, \pm 2$. The origin of the transverse field was attributed to a misalignment of $\theta_m = 1^o$ between the field direction and easy axis, extracted from the experiment of Friedman et al. [60] (see, also [4]). As it is noted by authors, this model is in "reasonably good agreement with all the experimental parameter values known so far". The differences with [51] are the following: 1) a more general spin-phonon interaction was considered; 2) transitions due to a transverse magnetic field were included; 3) longitudinal fields were not limited to the resonance near $H_z=0$. All that allowed them to get an independent description for each resonance in the experimental curve of Figure 5. To obtain a continuous description of the relaxation time vs $H_z$ the authors applied the Kirchhoff's rules by associating with each independent path from $|-10\rangle$ to $|10\rangle$ the probability current $J_n = d\rho_n/dt$, where $\rho_n$ is the reduced density matrix $\rho(t)$ for particular path n. This theory can effectively be applied to $Mn_{12}$, but after some correction given below. [52] allows to recover the observed Lorentzian line-shape for the resonance peaks, but the procedure that was used to get the experimentally observed line-width was "confusing". The authors cut off the calculated Lorentzians to the appropriate values since these last (the Lorentzians) were found extremely high and narrow. This procedure, of course, was supported by mathematical arguments, but the problem is that the authors have missed (as well as Fort et al. in [51]) in their theory most important contribution, namely, the value of the tunneling splitting in denominator of their



formula for the tunneling rate. As we will discuss below, exactly this missed term gives most important contribution to the width of the resonance line.

It was also shown in [52], that even and odd resonances should have different size since even resonances are induced by $S_+^4$ or $S_-^4$, whereas odd resonances are induced by combinations like $S_x S_+^4$ or $S_x S_-^4$ (similar effect was found experimentally by Thomas and Barbara [11,21]). In addition, based on experiments of Caneschi et al. in dilute samples [70] (compared with the same measurements on powder sample), authors completely ignored any dipolar fields as well as any hyperfine couplings since its contribution should give rise to the Gaussian distribution, whereas the peaks of relaxation rate, observed experimentally, have the Lorentzian shape. However, we have noted already (see the next to last paragraph before Section 4.2.), that it is not so easy to conclude from the present experiments (at $T<T_B\sim 3$ K) about the real shape of these peaks (even if they are the Lorentzians) close to tails. Even if the phonons dominate in the relaxation mechanism at $T>2$ K (with no doubts), the influence of dipolar and hyperfine couplings can not be negligible because (at least) the relaxation behaves non-exponentially at these temperatures (see [21,53,57-59,62] and references therein).

In the following we would like to give a qualitative picture of the Thermally Assisted Tunneling phenomenon including couplings to the environment (with respect to the resonance line-width). We do not state that our calculations are complete and want just to show how this works. (Note, that the similar calculations have been done by Prokof'ev and Stamp several years ago but they did not publish it)

The Hamiltonian of interest for the $Mn_{12}$ molecule can be written as follows:

$$H=H_G + H_{sp\text{-}ph} + H_{hyp} + H_{dip}, \qquad (21)$$

where $H_G$ is given by Eq. (1) and we take spin-phonon Hamiltonian in the simplest form of Eq. (19) with $V(\mathbf{q})=D(S_x S_z+S_z S_x)=(D/2)[(S_+ + S_-)S_z + S_z(S_+ + S_-)]$. The third term in Eq.



(21) describes the hyperfine interaction of central spin $\mathbf{S}$ with nuclear spins $\boldsymbol{\sigma}_k$ of each Mn ion (i.e., N=12 nuclear spins):

$$H_{hyp}=1/S\sum_{k=1}^{12}\ (\eta\,\omega_k/2)\,\mathbf{S}\boldsymbol{\sigma}_k. \tag{22}$$

And, finally, the last term gives the dipolar intermolecular interaction:

$$H_{dip}=1/2\sum_{i\neq j}\ V_{i,j}\mathbf{S}_i\mathbf{S}_j, \tag{23}$$

where $V_{i,j}$ depends on cube of the inverse distance between molecules. Let us assume now that the system is close to the characteristic magnetic field $H_n$ (Eq. (3)) when some pairs of levels from the two opposite sides of the barrier come into resonance. As this pair of levels is well separated from the others, we can truncate the giant spin Hamiltonian $H_G$ to a two-level one $H_{m,n-m}$ and consider the pair of levels $|m\rangle$ and $|n-m\rangle$ with the corresponding energies of $E_m^0$ and $E_{n-m}^0$ ($E_m^0= -Dm^2-K_{\parallel}m^4$):

$$H_{m,n-m}=\Delta_{m,n-m}\tau_x + \xi_{m,n-m}\tau_z, \tag{24}$$

where $\Delta_{m,n-m}$ is the tunneling splitting (due to fourth-order anisotropy terms, transverse components of dipolar and hyperfine interactions and transverse components of external magnetic field which can be originated in the experimental misalignment of the crystal), $\xi_{m,n-m}$ is longitudinal bias:

$$\xi_{m,n-m}=(2m-n)[(E_m^0-E_{n-m}^0)/(2m-n)+1/S\sum_{k=1}^{12}\ (\eta\,\omega_k/2)\sigma_k^z+\sum_{j\neq0}\ (V_{0,j}/2)m_j^* -g\mu_B H_z]/2 \tag{25}$$

and $\tau_x$, $\tau_z$ are the Pauli matrixes. In Eq. (25) the third term comes from the longitudinal part of dipolar interaction, where $m_j^*$ is the spin state of j-th molecule. In this case we can get for the maximum value of the tunneling probability with no phonons:

$$P_{m,n-m}^{(0)}=\Delta_{m,n-m}^2/(\xi_{m,n-m}^2 + \Delta_{m,n-m}^2) \tag{26}$$

The longitudinal hyperfine couplings give a Gaussian spread to each giant spin energy level $E_m$. This means that each energy level $E_m$ is actually split into a Gaussian multiplet



with N+1 different polarization groups of N=12 nuclear spins. In reality one should consider tunneling between sublevels from opposite sides of the barrier, which are in the resonance for a given field. However, in the simplest case (zero approximation) we will consider the longitudinal internal bias (which comes from the hyperfine coupling) just as some "mute" variable $\varepsilon$ with the Gaussian distribution:

$$G(\varepsilon)=(2\pi\sigma_o^2)^{-1/2}\exp(-\varepsilon^2/2\sigma_o^2). \tag{27}$$

In this case the half-width $\sigma_o \sim N^{1/2}\omega_0$ with $\omega_0=\langle\omega_k\rangle$ and according to the experimental results given above [53,21,87] $\sigma_o\approx 6$ mT (i.e., $\omega_0\approx 1.75$ mT). As for the dipolar coupling, we do not know it reliably (except from some preliminary measurements which give $E_D=2\sigma_o\sim 20$ mT) and, therefore, for the distribution of dipolar fields we take the same value $\sigma_o\approx 6$ mT. A remark should be made here. The distribution of dipolar fields strongly depends on the value of the magnetization (e.g. $M\approx 0$ after zero-field cooling or $M\approx M_s$ after field cooling, where $M_s$ is the saturated magnetization) and on the shape of the sample. Only at sufficiently strong annealing $M < 0.5\ |M_s|$ one can observe the Gaussian distribution for dipolar fields. In the following we assume zero-field cooling (an annealed sample). Thus, we can rewrite Eq. (25) in a more transparent form ($\varepsilon$ includes the contributions from both hyperfine and dipolar fields):

$$\xi_{m,n-m}=(2m-n)[g\mu_B(H_n-H_z)+\varepsilon]/2, \tag{28}$$

where $H_n$ is the characteristic field from Eq. (3) (actually $H_n$ are calculated using Eq. (1)). Choosing a Gaussian distributions for both dipolar and hyperfine fields, and with the same $\sigma_o$, is a great simplification which leads to think that dipolar fields are more or less ignored, since they behave as "non-interacting" hyperfine fields. Strictly speaking, we are not far from this suggestion in our simple model, but here we do not see the reasons to play with the form of distributions. In reality, however, one should include also the flip-flop transitions between molecules which come from the transverse part of



dipolar interaction. Moreover, when system relaxes, the total bias field (internal plus external) changes in time. The flipping of the molecules produces time-dependent fluctuations of dipolar fields which also cause the transitions in the nuclear subsystem. Therefore, each molecule feels rapidly fluctuating field $\xi = \xi(t)$ [44,47,61,72,73]. This means that the Boltzmann distribution is time-dependent and an additional time-dependence produces deviations from the exponential law (and these deviations are increasing with decreasing of the temperature). Ignoring all that, we assume here that all these processes are fast enough to keep the thermal quasi-equilibrium at each side of the barrier with the "static" Boltzmann distribution. This leads, of course, to the exponential relaxation, but the above simplification is sufficient for our purpose which consists in estimating the width of the resonance line in the case of annealed sample.

The transverse part of the hyperfine interactions contributes to the Berry phase of a central spin in producing a random complex phase (for a details see Tupitsyn et al. [23,31] and Prokof'ev and Stamp [47,72,73]). To our present concern, one can say that the transverse hyperfine interaction will act as a transverse field, and therefore change the tunnel splitting to an effective one:

$$\Delta_{m,n-m} = \Delta_{m,n-m}(\Phi), \tag{29}$$

where $\Delta_{m,n-m}(0)$ is the tunneling splitting in zero external transverse magnetic field with no hyperfine interactions and $\Phi$ is an additional phase from hyperfine interactions. It would be wrong to say that the distributions of internal transverse and longitudinal fields are always the same. However, in zero approximation (and for strongly annealed sample) we can use this suggestion and consider the Gaussian distributions with the same width $\sigma_0$ for both transverse and longitudinal fields (of course, it is easy to take different $\sigma_0$ for these fields).



Next, we should take into account the magneto-acoustic interaction. It was shown by Kagan and Maksimov [74] that the correct contribution (to all orders in $\Delta$) from inelastic phonon processes to the transition rate can be written as follows:

$$\Gamma_{m,n-m} = 2\Delta_{m,n-m}^2 W_m / (\xi_{m,n-m}^2 + \Delta_{m,n-m}^2 + \eta^2 W_m^2), \qquad (30)$$

where $\eta W_m$ is the phonon line-broadening. Note that this equation clearly shows the expected Lorentzian form vs a longitudinal fields for a system at quasi-equilibrium (see Eq. (28)). The most important transition here is the transition to the next upper level (from $|m\rangle$ to $|m-1\rangle$) with phonon absorption. From Eq. (19) with $V(\mathbf{q}) = D(S_x S_z + S_z S_x)$ we get (see also [63-65,90]):

$$W_m^{(1)} = (3D^2(S+m)(S-m+1)(2m-1)^2 E_{m-1,m}^3)(8\pi\rho c^5 \eta^4 [\exp(E_{m-1,m}/T)-1])^{-1}, \qquad (31)$$

where $E_{m-1,m} = E_{m-1} - E_m$, $\rho = M/a^3$ is the mass density (a is the lattice constant), $c = (k_B/\eta)\Theta_D(V_o/6\pi^2)^{1/3}$ (see [75]) is the sound velocity with the Debye temperature $\Theta_D$ and $V_o$ is the unit cell volume. According to recent measurements of specific heat in $Mn_{12}$ by Gomes et al. [76] $\Theta_D = (38 \pm 4)$ K and from [1], we can take $V_o = 3716$ A$^3$ and $\rho = 1.83 \times 10^3$ kg/m$^3$.

To see the dominant contribution to the Lorentzian line-width (see Eq. (30)), one must compare $\xi_{m,n-m}$, $\Delta_{m,n-m}$ and $\eta W_m$. As already noted, the estimated width of the Gaussian distribution of internal fields is approximately $2\sigma_0 = 12$ mT (for hyperfine interaction, at least). Let us concentrate for the moment to the high temperature regime, with thermally activated tunneling from levels m smaller or equal to 4. In zero field $\Delta_{m,-m}(0)$ gives $1.1 \times 10^{-2}$ K and 0.34 K, for m=4 and m=2 respectively, whereas $\eta W_4^{(1)} \approx 1.02 \times 10^{-5}$ K and $\eta W_2^{(1)} \approx 6.08 \times 10^{-7}$ K (for T=2.6 K and c=1.4×10$^3$ m/s). It is only in the region of m=6 or larger the values of $\eta W_m^{(1)}$ and $\Delta_{m,-m}(0)$ become more or less



comparable. In order to take into account odd values of m (as well as nonzero n), we have to include internal transverse fields (as it is discussed above Eq. (29)).

We can also consider the phonon-assisted transitions with $\delta m = \pm 2$. In this case we take for simplicity $V(\mathbf{q}) = D(S_x^2 - S_y^2) = D(S_+^2 + S_-^2)/2$. For transition from $|m\rangle$ to $|m-2\rangle$, this yields (see also [52]):

$$W_m^{(2)} = (3D^2(S+m)(S+m-1)(S-m+2)(S-m+1)E_{m-2,m}^3)(8\pi\rho c^5 \eta^4 [\exp(E_{m-2,m}/T)-1])^{-1}. \quad (32)$$

This equation gives $\eta W_4^{(2)} \approx 2.99 \times 10^{-5}$ K and $\eta W_2^{(2)} \approx 1.16 \times 10^{-5}$ K (also for T=2.6 K). These numbers show that transitions with $\delta m = \pm 2$ can not change the situation with relative contributions of the values $\eta W_n$ and $\Delta_{m,-m}(0)$ to the Lorentzian line-width, however, we will use the average value $W_m = (W_m^{(1)} + W_m^{(2)})/2$ in Eq. (30). Of course, spin-phonon interaction contains other terms but it is unlikely that they will give contribution which is of the order of magnitude larger than $W_m$.

This means that in thermally activated regime (near the top of the barrier) the line-width is defined mainly by the tunneling splitting of the resonant levels and by the internal longitudinal fields (see Eq. (28)), but not by the phonon line-broadening. Note, however, that the phonons play an essential role in "linking" the states on the same side of the barrier (otherwise only the ground-state would be occupied, unless one admits that *all* the bias is dynamical). The dipolar flip-flop processes also can cause transitions between the energy levels providing thermal equilibrium.

We should note that Villain and Fort et al. in [51] as well as Leuenberger et al. in [52] have slightly different definitions for the tunneling rate, but which in general, look like:

$$\Gamma_{m,k} = 2\Delta_{m,k}^2 W_m / (\xi_{m,k}^2 + \eta^2 W_m^2), \quad (33)$$

where m and k are the levels in the resonance and $\xi_{m,k}$ is the bias of Eq. (25) (or Eq. (28)) with no internal bias fields. According to this formula, the half-width of the



resonance peak should be of the order of $\eta W_m$ which is actually very small in comparison with $\Delta_{m,-m}$ in zero field for upper levels where the tunneling takes place at T>2 K. (see above, just after Eq. (32)). The reason of this problem, is that these authors missed the term relative to the intrinsic width $\Delta$ in their tunneling rate (see Eq. (30) and (33)) which is absolutely not negligible for thermally activated tunneling.

In the thermally activated regime, the value of the tunneling rate can be evaluated for each resonance (i.e. each value of n) by taking the product of the intra-well (Boltzmann) and inter-wells (tunneling) transition probabilities. This has to be summed up to all the contributions from different $|m\rangle$ :

$$\tau^{-1}_n(H_z)=Z^{-1}(H_z)\sum_m \Gamma_{m,n-m}\exp[(-E_m^{\ 0}-g\mu_B H_z m)/k_B T], \qquad (34)$$

where $Z(H_z)$ is the partition function. (For simplicity we omit here the time dependence of the Boltzmann factor, i.e. we still assume quasi-equilibrium). The plot of $\tau^{-1}_n(H_z)$ gives the expected Lorentzian line-shape of resonance peaks. The width is rather sensitive to parameters such as the distribution of internal fields or the sound velocity. However realistic values for these parameters allow to obtain the experimentally measured line-widths. To give some numbers, we obtained the width of about 25 mT (see [60]) with $\sigma_o = 6$ mT at T=2.6 K. To get the same height, it was necessary to change only one parameter from all the set. We put the sound velocity to the value of $c=1.4x10^3$ m/s, the other parameters are the same as given above (see Eq. (1) and Eq. (31)). Our curves show slightly faster decay near the tails (which is the consequence of the averaging over a Gaussian distribution of fluctuating internal fields). However, if we neglected these internal fields, Eq. (34) would give too narrow and sharp resonance lines.



Before concluding, a few remarks should be done. We limited ourselves by taking an average of Eq. (30) over fluctuating internal fields (longitudinal in bias $\xi_{m,n-m}$ and transverse in $\Delta_{m,n-m}$) but this is just zero approximation, to show that even this approximation can lead to correct resonance line-width. As we noted already, in reality the interaction with the nuclear subsystem spreads each giant spin energy level into a Gaussian multiplet and one should consider all possible transitions between the resonant levels from the opposite sides of the barrier inside such multiplets. Moreover, the internal bias field $\varepsilon(t)$ is actually time-dependent (i.e., it varies not only in space) and produces, in general, a time-dependent Boltzmann factor $(Z^{-1}(H_t(t)) \bullet (\exp[(-E_m^0 + mH_t(t))/k_BT]$ where $H_t(t) = g\mu_B(H_z - \varepsilon(t))$ and, as a consequence, deviation from exponential relaxation occurs.

To conclude our discussion of the high-T behavior of the relaxation in $Mn_{12}$-acetate, we should emphasize the following main results: 1) At relatively high temperatures (T>2 K) the relaxation is dominated by the Phonon-Assisted Tunneling Mechanism; 2) The Lorentzian shape of the resonance peaks is determined by Eq. (30), describing the tunneling between the levels which are in resonance in a given magnetic field; 3) Despite the dominant role of phonons the resonance line-width is determined mainly by the value of the tunneling splitting $\Delta_{m,n-m}$ together with internal bias $\varepsilon$ (Eq. (28)) which originates from the hyperfine and dipolar interactions; 4) All mentioned theories predict exponential relaxation of the magnetization while the experiments show a non-exponential behavior at these temperatures (see [4,6,7,57-59,21,53,62]). This fact clearly shows the important role played by intermolecular interactions as well as by interactions with nuclear spins which both produce time-dependent bias field causing the non-exponential behavior of relaxation.



### 4.3. Ground-state Tunneling.

Let us now concentrate on the low-temperature limit of $T<<T_c$ where tunneling takes place on the ground-state only. In the instanton approach (see below), we assume that $k_B T<<\Omega_0$, where $\Omega_0$ is the "bounce frequency" of the instanton transition (which is roughly the distance from the ground-state level to the first excited one. In this case only the lowest levels are thermally populated and we can truncate the "giant spin" Hamiltonian (Eq. (1) or Eq. (12)) to an effective one (which is valid only at energies $<<\Omega_0$), describing two states $|S_z\rangle = \pm S$ and their mixture, separated by the tunneling splitting $\Delta<<\Omega_0$. Moreover, we should couple the "giant spin" $\mathbf{S}$ of each molecule to the nuclear "spin-bath" $\{\sigma_k\}$ with k=1,2,...,N. In the case of hyperfine coupling, the latter can be described by Eq. (22) with $\omega_k<<\Omega_0$ ($\omega_k>>\Delta$ in most cases) and with N equal to actual number of nuclear spins inside molecule (one can take N=12 for $Mn_{12}$ neglecting the effect of hydrogen…, and N=146 for $Fe_8$, including 120 hydrogen, 18 nitrogen and 8 bromine atoms, neglecting the few percents of Fe57). Without hyperfine interactions, the total nuclear spectrum, containing $2^N$ states, is almost completely degenerated, with only a tiny spreading $\sim T_2^{-1}$ of levels caused by the inter-nuclear dipolar interactions. With the hyperfine interaction, the nuclear levels spread into a Gaussian multiplet of N+1 polarization groups around each giant spin level (see [47,61,72,44,23]). The half-width of this distribution $\sigma_0$ is of the order of $\omega_0 N^{1/2}$ (where $\omega_0 = (1/N) \sum_k^N \omega_k$). Note that, in each polarization group, the hyperfine levels are also distributed according to a Gaussian half-width $\sim T_2^{-1}$. Typically the different polarization groups completely overlap within of the Gaussian envelope of Eq. (27). The nuclear $T_2$-processes (transverse relaxation) are responsible for transitions inside each polarization group, whereas the nuclear $T_1$-processes (longitudinal relaxation) provide transitions between different polarization groups. According to [45], the half-width of the hyperfine



distribution in Fe$_8$ is of the order of 0.6 mT, yielding $\omega_0 \sim 0.05$ mT (for the estimate we take only hydrogen atoms with $\sigma=1/2$, i.e., N=120). Since $\omega_0 >> T_2^{-1}$ (typically $T_2^{-1} \sim 10^{-7}$ K), the nuclear spin dynamics is slaved by dynamics of **S.** We also suggest that the $T_1$-processes are long and therefore not relevant, at these temperatures. The reason is that the $T_1$-processes are driven by the dipolar flip-flop transitions, which are essentially frozen at $T << T_c$. In general, we should consider two effects: (i) the effect of nuclear spins on the giant spin dynamics during the tunneling and (ii) the effect of the motion of **S** on the nuclear spins. Both effects have to be handled in a self-consistent way.

Using the instanton technique, we can write the effective Hamiltonian for a single molecule in the following way [23]:

$$H_{eff}{}^{(1)} = [2\Delta_0 \tau . \cos(\pi S - \beta_0 \mathbf{n \cdot H} + \sum_{k}^{N} \alpha_k \mathbf{n \cdot \sigma_k}) + H.c.]$$

$$+ (1/2)[\tau_z \sum_{k}^{N} \omega_k{}^{\parallel} \mathbf{l_k \cdot \sigma_k} + \sum_{k}^{N} \omega_k{}^{\perp} \mathbf{m_k \cdot \sigma_k}]$$

$$+ \sum_{k \neq l}^{N} V_{kl}{}^{\alpha\beta} \sigma_k{}^{\alpha} \sigma_l{}^{\beta}, \qquad (35)$$

where $\boldsymbol{\tau}$ describes the "giant spin" of molecule ($\boldsymbol{\tau}$ and $\boldsymbol{\sigma}$ are both the Pauli matrices). The first term in Eq. (35) is non-diagonal term (due to $\tau_{\pm}$) which operates during transition of **S.** It produces a time-dependent field $\boldsymbol{\gamma}_k = (\omega_k \mathbf{S}/2S)$ acting on each $\boldsymbol{\sigma}_k$, and this leads to $\boldsymbol{\sigma}_k$ to flip. If we expand out the cosines, we see that we have a whole series of terms like $\sim \tau_{\pm} \Gamma_{\alpha\beta\gamma\delta} \sigma_{k1}{}^{\alpha} \sigma_{k2}{}^{\beta} \sigma_{k3}{}^{\gamma} \sigma_{k4}{}^{\delta} \ldots$ in which an instanton flip of the giant spin couples to many different nuclear spins simultaneously, i.e., a single instanton can simulate multiple transitions in the nuclear bath. The probability that $\boldsymbol{\sigma}_k$ will flip during a single instanton passage between two quasiclassical minima $|\mathbf{S}_1\rangle$ and $|\mathbf{S}_2\rangle$ is $|\alpha_k|^2/2$.



Thus, the average number of nuclear spins that will flip each time when **S** flips (so-called "co-flipping" amplitude) is approximately:

$$\lambda = (1/2) \sum_{k}^{N} |\alpha_k|^2 \qquad (36)$$

which, in principle, can be larger than 1. One can easily calculate the dimensionless constants $\alpha_k$ and $\beta_0$ in the case of the simple bi-axial Hamiltonian of Eq. (6) with $H_{hyp}$ given by Eq. (22). The answer is [23,31]:

$$\alpha_k \mathbf{n} \cdot \boldsymbol{\sigma}_k \approx (\pi \omega_k / 2\Omega_0)[-i\sigma_y + (D/(E+D))^{1/2}\sigma_x] \qquad (37)$$

$$\beta_0 \mathbf{n} \cdot \mathbf{H} \approx (\pi g \mu_B S / \Omega_0)[-iH_y + (D/(E+D))^{1/2}H_x], \qquad (38)$$

where **n** is a unit vector in the (**xy**) plane and the "bounce frequency" $\Omega_0$ reads as:

$$\Omega_0 \approx 2S(ED)^{1/2}. \qquad (39)$$

(For non-zero values of $K_\perp$ see Eq. (14-18)). Eq. (37) tells us that $\lambda < 1$ in the particular case of E and D as in $Fe_8$ (or $Mn_{12}$, note that non-diagonal fourth-order terms renormalize Eq. (37,38) but for $K_\perp/k_B = -3.28 \times 10^{-5}$ K it gives $\alpha_k \sim (\omega_k/\Omega_0 C_\perp)$ with $C_\perp \approx 1.56$).

The second term in Eq. (35) is diagonal, which operates when S is in one from its two qusiclassical minima. Let us introduce two corresponding fields $\gamma_k^{(1)}$ and $\gamma_k^{(2)}$. In general case (any non-zero external and internal magnetic field) $\mathbf{S}_1$ and $\mathbf{S}_2$ are not antiparallel. It is easy to see that the sum and the difference between these two vectors define $\omega_k^{\parallel}$ and $\omega_k^{\perp}$ i.e.,

$$\omega_k^{\parallel} \mathbf{l}_k = \gamma_k^{(1)} - \gamma_k^{(2)} \qquad (40)$$

$$\omega_k^{\perp} \mathbf{m}_k = \gamma_k^{(1)} + \gamma_k^{(2)}, \qquad (41)$$

where $\mathbf{l}_k$ and $\mathbf{m}_k$ are mutually perpendicular unit vectors. The longitudinal coupling $\omega_k^{\parallel}$ gives the change in energy of $\sigma_k$ before and after **S** flips (i.e., the difference in effective



field acting on $\sigma_k$ before and after transition of **S**). The transverse coupling $\omega_k^{\perp}$ defines the deviation of initial and final orientations of **S** from the $\pm$**z**-direction (which is easy-axis for the Hamiltonian of Eq. (12)). In the case of biaxial Hamiltonian for small values of $H_x$ ($g\mu_B H_x << S(E+D)$) we get:

$$\omega_k^{\parallel} \mathbf{l}_k \cdot \sigma_k \sim \omega_k \cdot \sigma_z \tag{42}$$

$$\omega_k^{\perp} \mathbf{m}_k \cdot \sigma_k \sim (\omega_k g \mu_B H_x / (2S(E+D))) \cdot \sigma_x \tag{43}$$

Finally, the third term in Eq. (35) describes very weak internuclear dipolar coupling ($|V_{kl}^{\alpha\beta}| \sim T_2^{-1}$). To complete our Hamiltonian, we should also include the dipolar-dipolar interactions between molecules. Ignoring transverse part of this interaction (which leads to flip-flop processes), we get:

$$H_D = (1/2) \sum_{\mu \neq \nu} V_{\mu,\nu}^{(D)} \tau_z^{\mu} \cdot \tau_z^{\nu}, \tag{44}$$

where $|V_{\mu,\nu}^{(D)}| \sim 1$ mK. Thus, to work in the low-T Quantum Regime, one may use the following effective Hamiltonian:

$$H_{eff} = H_{eff}^{(1)} + H_D. \tag{45}$$

We do not include the spin-phonon interaction into this Hamiltonian since the phonons can play no role at $T < T_c$.

Now, after the effective low-T Hamiltonian is established, we would like to discuss briefly some effects which follow from the coupling of the central spin to the spin bath (for detailed explanation one should read original papers [72,47] and review [73,61]).

Let us start from the nondiagonal term in Eq. (35) (the first term). Since $\alpha_k$ is complex quantity, one can expect two effects. The imaginary part of $\alpha_k$ gives a renormalization of the effective tunneling splitting depending on the coupling constants



$\omega_k$. One can expect an increase (decrease) of the effective tunneling splitting with $\omega_k$. The width of the distribution of hyperfine fields also depends on $\omega_k$. It should also increase with $\omega_k$. The real part of $\alpha_k$ adds an extra random phase to the Haldane phase $\beta_0 \mathbf{n} \cdot \mathbf{H}$. As we have seen already, the average number of nuclear spins that flip together with S is proportional to $|\alpha_k|^2$. These flips modify the total phase of the bath state and, consequently, randomize the phase of the giant spin (between the instanton and anti-instanton) producing phase decoherence of the tunneling process that can completely block the latter (this is called, "*topological decoherence*"). Indeed, we know that a half-integer spin can not tunnel in zero transverse field. We can imagine that in different molecules the transition of **S** is accompanied by the different number of the nuclear spins. If the total flipping spin (**S** together with the nuclear spins $\boldsymbol{\sigma}_k$) is integer, the tunneling is allowed. Otherwise, the tunneling is blocked. Thus, the possibility to tunnel depends on the particular environment state inside of each molecule (nuclear spins), and over the entire sample the transition of **S** can happen at random.

The second term in Eq. (35) produces an internal bias field (we put here $\omega_k^\perp = 0$) $\varepsilon = (1/2) \sum_k \omega_k^\parallel \sigma_k^z$ acting on **S**. Together with the first term it gives the Hamiltonian of the biased two-level system (see Eq. (24)) with an effective tunneling splitting $2\Delta_\Phi = 2\Delta_0 \cos(\Phi)$ where $\Phi$ is the complex phase (we assume here that there is no nuclear spin dynamics itself, i.e., $|V_{kl}^{\alpha\beta}| = 0$). In this case the tunneling probability is given by:

$$P^{(0)}(t) = (4\Delta_\Phi^2/E^2)\sin^2(Et), \qquad (46)$$

$$E = \pm(\varepsilon^2 + 4\Delta_\Phi^2)^{1/2}. \qquad (47)$$

Since $\varepsilon \gg \Delta_{-10,10}$ (in our case $\omega_0 \gg \Delta_{-10,10}$), only a small fraction of molecules is not pushed away from the resonance by the additional longitudinal field $\varepsilon$ ($\varepsilon$ depends, of course, on the particular environment state) and, therefore, are able to tunnel. To



estimate this small number of molecules that are close to the resonance, we should take the average of $P^{(0)}(t)$ over the ensemble of the molecules with different $\varepsilon$ weighted by the Gaussian distribution of Eq. (27). This gives (in what follows, for nuclear spins we use notation $\xi_0$ instead of $\sigma_0$) for $\xi_0 >> \Delta_0$ (see Chapter 4 in [72]):

$$P^{(0)}(t) \sim \rho \sum_{k=0}^{\infty} J_{2k+1}(4\Delta_\Phi t), \qquad (48)$$

which oscillates as the Bessel function $J_{2k+1}(z)$ with the amplitude equal to:

$$\rho = (2\pi)^{1/2} \Delta_\Phi / \xi_0. \qquad (49)$$

The value of $\rho$ estimates the fraction of molecules that are able to tunnel. It is easy to see that $\rho << 1$. This mechanism (called "*degeneracy blocking*") very effectively can block the tunneling. However, if we include the dynamics of the nuclear bath ($|V_{kl}^{\alpha\beta}| \neq 0$), the situation changes dramatically. Due to the interaction between the nuclear spins, bias energy $\varepsilon$ becomes time-dependent: within each polarization group $\varepsilon(t) = \varepsilon + \delta\varepsilon(t)$ passes over all the energy range $\sim T_2^{-1}$ and this gives to the system a resonance window. Inside this window the total bias field fulfills the condition $\xi = \xi_H + \varepsilon(t) < \Delta$ (where $\xi_H$ is external bias field) and central spin can tunnel. (The same mechanism can also destroy coherence by pushing molecules away from the resonance window.) Note that tunneling can take place only between the polarization states M and –M (M=$S_z$) because the energy of the final state $E_f$ should be in the resonance with the energy of the initial state $E_i$ ($|E_f-E_i|$ should not exceed $\Delta_0$, at least). This means that if initial polarization state of molecule is M, 2M nuclear spins flip when **S**. As we discussed above, the *average* number of nuclear spins that will flip together with **S** is $\lambda << 1$ (for Fe$_8$ and Mn$_{12}$). In fact, better possibility to tunnel is realized for molecules with initial polarization state M=0. Molecules with $M_{in} \neq 0$ also can tunnel (with $\delta M=2M$) but the contribution into the statistics from these events falls very rapidly with



the increase of M. Actually $\Delta_\Phi(M)$ falls as $\sim(\lambda^M/M!)^{1/2}$ for $M \gg \lambda$ (see Chapter 4 in [47]).

If $\omega_k^\perp \neq 0$, there is a transverse magnetic field acting on environmental spins. Due to this field the initial and final directions of the nuclear spins are not parallel (or antiparallel) to each other. Assume that all nuclear spins are initially aligned in $\boldsymbol{\gamma}^{(1)}$ (before **S** flips). After **S** flips, nuclear spins $\boldsymbol{\sigma}_k$ are not parallel (antiparallel) to the new field $\boldsymbol{\gamma}^{(2)}$ acting on them (note that the instanton flip of **S** is a sudden perturbation for nuclear spins, who undergo a non-adiabatic transition)**.** This new state is not an eigenstate of the Hamiltonian and nuclear spins must relax in making transitions to avoid misalignment with $\boldsymbol{\gamma}^{(2)}$ (this transitions transform their wave-function to the exact eigenstate). Thus, the tunneling of **S** can be suppressed (depending on how slow are the transitions in the particular environment state) since the initial and the final states of the nuclear bath are not exactly orthogonal. This mechanism is known as "*orthogonality blocking*" mechanism (see Chapter 4 in [72,47]).

All these effects can be handled by three different kinds of the averaging procedure and the final answer can be obtained by combining them depending on its importance in each particular case (for analytical expressions see [72,47,73,61]). We have seen also that dynamics of the nuclear bath is extremely important since, producing rapidly fluctuating hyperfine field, it can help the system to find a resonance window by "scanning" over all the range of the bias energy (which for single molecule is of the order of $T_2^{-1}$). For the particular case of $Fe_8$ or $Mn_{12}$ we have $\lambda \ll 1$ which means that, in general, molecules in the resonance window relax incoherently (only molecules in nuclear polarization state with M=0 can relax coherently in this case). The relaxation rate for such incoherent process is given by [47]:

$$\tau_N^{-1}(\xi) \approx \tau_0^{-1}\exp(-|\xi|/\xi_0), \qquad (50)$$



$$\tau_0^{-1} \approx 2\Delta_{-10,10}{}^2/\pi^{1/2}\,\xi_0. \tag{51}$$

However, these equations describe just the initial stage of the relaxation. Continuous relaxation requires to bring more and more molecules in resonance that can be provided by fluctuations of the dipolar field across the sample [44]. When the spin of a given molecule flips, it produces the time-dependent magnetic field (long-range) which can push some molecules away from the resonance and bring other ones into the resonance window depending on the nuclear bath state of each molecule.

In order to investigate the problem of the relaxation in quantum regime (ground-state tunneling), Prokof'ev and Stamp in [44] introduced a kinetic equation for distribution function $P_\alpha(\xi,\mathbf{r},t)$ which gives the probability to find a molecule at position $\mathbf{r}$ with polarization $\alpha=\pm 1$ (i.e., $|S_z\rangle=\pm S$) having a bias energy $\xi$ at time t. This equation reads as:

$$dP_\alpha(\xi,\mathbf{r})/dt = -\tau_N^{-1}(\xi)[P_\alpha(\xi,\mathbf{r})-P_{-\alpha}(\xi,\mathbf{r})] - \sum_{\alpha'} \int d\mathbf{r}'\Omega_0^{-1} \int d\xi'\,\tau_N^{-1}(\xi')$$

$$\times[P_{\alpha\alpha'}{}^{(2)}(\xi,\xi',\mathbf{r},\mathbf{r}')-P_{\alpha\alpha'}{}^{(2)}(\xi-\alpha\alpha'V^{(D)}(\mathbf{r}-\mathbf{r}'),\xi',\mathbf{r},\mathbf{r}')], \tag{52}$$

where $P_{\alpha\alpha'}{}^{(2)}(\xi,\xi',\mathbf{r},\mathbf{r}')$ is a two-molecule distribution which gives the probability to find a second molecule with polarization $\alpha'$ and bias $\xi'$ if the first molecule is with $\alpha$ and bias $\xi$, correspondingly. The quantity $\Omega_0$ is the volume of unit molecular cell, $V^{(D)}(\mathbf{r})$ is the longitudinal part of the dipole-dipole interaction (see Eq. (44)), and integration $\int d\mathbf{r}'$ is performed over the sample volume. The first term of this equation describes the local tunneling relaxation whereas the second (which is analogous to a collision integral) describes the influence of the dipolar field produced by the spin flip of a molecule at site $\mathbf{r}'$. The solution of Eq. (52) (analytical or numerical) gives the magnetization M(t) as a function of time by the following obvious equation:

$$M(t) = \int d\xi' \int d\mathbf{r}'\Omega_0^{-1}[P_+(\xi,\mathbf{r})-P_-(\xi,\mathbf{r})]. \tag{53}$$



If at t=0 the sample is fully polarized, at short times the solution of Eq. (52) can be found analytically (for ellipsoidal shape, at least). At the beginning of the relaxation the number of the flipped molecules is small ($M(t)/M_s \ll 1$, where $M_s$ is the saturated magnetization) and, according to Anderson (see [55]), the field distribution of the randomly placed *dilute* static dipoles can be described by the Lorentzian:

$$P_\alpha(\xi) = [(1+\alpha M(t))/2] \cdot [(\Gamma_d(t)/\pi)/\{(\xi-\alpha E(t))^2 + \Gamma_d^2(t)\}]; \qquad (54)$$

$$\Gamma_d(t) = (4\pi^2 E_D/3^{5/2})[1-M(t)]; \qquad (55)$$

$$E(t) = c V_D(1-M(t)), \qquad (56)$$

where c is the sample shape-dependent constant and $V_D$ is the strength of the dipolar interaction ($V^{(D)}(\mathbf{r}) = V_D \Omega_0 [1-3\cos(\theta)]/r^3$). In this case (short times, the Lorentzian distribution) the two-molecule distribution function becomes factorizable (i.e., $P^{(2)}(1,2) = P(1)P(2)$) and Eq. (52) gives the square-root relaxation law [44]:

$$M(t)/M_s = 1 - (\tau_{short}^{-1} t)^{1/2}, \qquad (57)$$

where $\tau_{short}^{-1}$ is the relaxation rate:

$$\tau_{short}^{-1} = \eta \Delta_{10,-10}^2 \, P(\xi_D)/\eta \qquad (58)$$

with a sample shape-dependent constant $\eta$ and the normalized distribution of dipolar fields in a sample is $P(\xi_D)$. When the number of the flipped molecules becomes large enough (larger than $10 \div 15$ %), the field distribution of the flipped spins becomes non-Lorentzian and Eq. (52) should be solved numerically since Eqs. (54-56) are no longer valid. This equation also can be solved numerically for nonsaturated sample (i.e., $M_{in}/M_s < 1$, where $M_{in}$ – initial magnetization) as well as for samples of different geometry with $M_{in} = M_s$. The latter was done in [44] by kinetic Monte Carlo simulations. The main result of these simulations is that the short-time relaxation still obeys the square-root law with a sample geometry-dependent constant $\eta$ (very recently this result



was confirmed by numerical calculations of Cuccoli et al. in [81]). As for the relaxation in nonsaturated sample, very interesting result was obtained experimentally by Wernsdorfer et al. in [45] in $Fe_8$ and by Chiorescu et al. in $Mn_{12}$ [85]. They found that the short-time square-root law is accurately obeyed for both saturated and non-saturated samples. Moreover, the square-root law for a strongly annealed sample or in the presence of strong transverse field is even more pronounced, as it can be seen from the comparison of the Figure 18 ($M_{in}=M_s$) with Figure 32 ($M_{in}=0$)! The theory of Prokof'ev and Stamp (as it is) does not predict the short-time square-root relaxation law for annealed samples. Among other possible suggestions, one could say that the distribution function $P_{\alpha\alpha'}^{(2)}(\xi,\xi',\mathbf{r},\mathbf{r'})$ is also factorizable if $M_{in}$ is small, and all the difference with the case of the saturated sample is in numerical constant $\eta$. Anyway, the square-root law for annealed sample was confirmed in [25] by the comparison of the tunneling rates extracted from the relaxation experiments using Eqs. (57-58) and from the experiments based on Landau-Zener model [82,68,69]. These experimental results require further theoretical investigation. One should nevertheless mention that this square root relaxation is necessarily a short-time regime and a cross-over to another relaxation regime must be observed at long times or/and high temperatures. Such a cross-over has been observed in $Mn_{12}$ for both M$\rightarrow$M$_s$ [57] and M$\rightarrow$0 [85]. The new regime is exponential and corresponds to phonons recovery.

As we understand now, at low T the short-time square-root relaxation law is explained by dynamic dipolar interactions (but, to find a resonance window at the beginning of the relaxation, the system needs dynamic hyperfine interactions). The influence of dipolar fields can also be investigated by measurements of the tunnel splitting $\Delta(H_x)$ versus a transverse magnetic field. In $Fe_8$ at low T, this quantity depends strongly (near the nodes) on the value of the initial magnetization (as it was found



experimentally in [83] using Landau-Zener method), i.e. it depends strongly on the strength of dipolar interactions between molecules (see Figure 33). In what it follows we would like to show how to analyze this dependence using very simple language [84].

The tunneling probability in the Landau-Zener model depends on the sweeping rate of the longitudinal field $H_z$ in the following way (see, for example, [82]):

$$P_{m,m'}=1-\exp(\pi\Delta_{m,m'}^2(\xi_\parallel,\xi_\perp)/\eta\upsilon), \qquad (59)$$

where $\upsilon=g\mu_B(mm')^{1/2}dH_z/dt$, $dH_z/dt$ is the constant sweeping rate and, $\Delta_{m,m'}$ depends on dipolar bias fields $(\xi_\parallel,\xi_\perp)$. At large sweeping rates $P_{m,m'}\approx\pi\Delta_{m,m'}^2/\eta\upsilon$. Thus, the average probability over the distribution of the dipolar fields reads as:

$$P_{m,m'}(t_0)\approx(\pi/\eta\upsilon)\int d\xi_\parallel\int d\xi_\perp G(\xi_\parallel,\xi_\perp,t_0)\,|\Delta_{m,m'}(\xi_\parallel,\xi_\perp)|^2, \qquad (60)$$

where $G(\xi_\parallel,\xi_\perp,t_0)$ is the distribution of the dipolar fields in a sample. We have calculated numerically $G(\xi_\parallel,\xi_\perp,t_0)$ for different sample geometries (sphere, cube, parallelepiped) and for different initial magnetizations ($M(t_0)=M_{in}$). In the case $M_{in}<<M_s$ (e.g. zero-field cooled sample) the distributions are the Gaussians along all the axes (**x,y,z**). Then, for $dH_z/dt>>1$ Eq. (60) can be simplified to (for m=S and m'=-S):

$$P_{S,-S}(t_0)\approx(\pi/\eta\upsilon)G_\parallel(H_z)\int d\xi_x\int d\xi_y\,|\Delta_{S,-S}(\xi_\parallel=H_z,\xi_x,\xi_y)|^2G_\perp(\xi_x,\xi_y); \qquad (61)$$

$$G_\perp(\xi_x,\xi_y)\approx(2\pi E_{DX}E_{DY})^{-1}\exp[-\{(\xi_x-H_x)^2/(2E_{DX}^2)+(\xi_y-H_y)^2/(2E_{DY}^2)\}]; \qquad (62)$$

$$G_\parallel(H_z)\approx(2\pi E_{DZ}^2)^{-1/2}\exp[H_z^2/2E_{DZ}^2], \qquad (63)$$

Note, that a molecule can tunnel only in the total bias field $\xi_\parallel H_z<\xi_0/g\mu_B S$. Using this condition we put (approximately) $\xi_\parallel\sim H_z$. Next we would like to calculate $P_{S,-S}(t_0)$ near the nodes of $\Delta_{S,-S}(H_x,H_y)$. Let us call $\boldsymbol{\gamma}$ the integral in Eq. (61). Eq. (35) gives the effective tunnel splitting:

$$\Delta_{S,-S}=\Delta_0\cosh[\pi H_y/T_y+i\pi H_x/T_x], \qquad (64)$$



where $T_x$ is the oscillation period along the **x**-axis and $T_y$ is the "related" period along the **y**-axis. For Eq. (12) we get:

$$T_x/T_y=[(D+E)/D]^{1/2}. \tag{65}$$

Near the node (the first node, for definiteness) we can put:

$$H_x=H_x^0+\delta H_x+H_x^{dip}, \quad H_y=H_y^0+\delta H_y+H_y^{dip}, \tag{66}$$

where $(H_x^0,H_y^0)$ is the position of the node in the $(H_x,H_y)$ plane, $\delta H_{x,y}$ is the distance from the node and $H_{x,y}^{dip}$ is the contribution from the dipolar fields $(\xi_{x,y})$. In the first node (m=S, m'=-S) $H_y^0=H_z=0$ and $H_x^0/T_x=\pi(n+1/2)$. If $\delta H_{x,y}$ is small, it is easy to calculate $\boldsymbol{\gamma}$ analytically:

$$\boldsymbol{\gamma}\approx\Delta_0^2[\Psi_x^2+\Psi_y^2+(\pi^2)(E_{DX}^2+E_{DY}^2)], \tag{67}$$

$$\Psi_x=\pi\delta H_x/T_x, \quad \Psi_y=\pi\delta H_y/T_y. \tag{68}$$

Thus, for $E_{DX}=E_{DY}=E_D$ (zero-field cooled after annealed at high temperature) exactly in the node ($\Psi_{x,y}=0$) we get:

$$\boldsymbol{\gamma}\approx2\pi^2\Delta_0^2E_D^2. \tag{69}$$

When $M_{in}$ decrease, the width of the dipolar fields distribution increases. With increasing of $E_D$ the value of $\boldsymbol{\gamma}$ in the nodes also increases. This result explains the experimental behavior of the tunneling splitting in the nodes (since $\Delta_{S,-S}$ is proportional to $\boldsymbol{\gamma}^{1/2}$). To check these formulas we have calculated $\boldsymbol{\gamma}$ numerically using Eq. (12). The Figure 34 shows that (i) for different values of $E_D$ (i.e., depending on annealing) $\boldsymbol{\gamma}$ in the nodes really behaves like $E_D^2$ (see Figure (c)); (ii) around the nodes $\boldsymbol{\gamma}$ has parabolic dependence on applied transverse magnetic field in accordance with Eq. (67) (see Figure (b)). The Figure 35 presents the same $\boldsymbol{\gamma}$ as in Figure 34 (a) but calculated with a small "misalignment" angle $\theta_m=1°$. The curves in Figure 35 behave more similar to the experimental ones with respect to relative value of $\boldsymbol{\gamma}$ in different nodes.

Above considerations concludes our discussion on the quantum regime of relaxation (low-T ground-state tunneling) in $Fe_8$ and $Mn_{12}$. Strictly speaking, all these effects are not restricted to these systems but are valid for all mesoscopic systems where quantum tunneling is associated with extremely small tunneling splitting, i.e. $\Delta \approx 10^{-n}$ with n much larger than one ($n \approx 7$ and $n \approx 11$ in $Fe_8$ and $Mn_{12}$). The number of phonons available at these temperatures being negligible, phonons are not really relevant. On the contrary, other types of fluctuations are numerous in this regime and this is the case for fluctuations of the spin bath. The main result, which we would like to emphasize here, looks very simple and straightforward. All the physics in this limit depends on hyperfine and dipolar interactions. Any particular result is just the consequence of the first one. Finally we would like to say that the molecules $Fe_8$ and $Mn_{12}$ discussed in this paper are ferrimagnetic with a large non-compensated moment, i.e. their physics is dominated by the ferromagnetic order parameter. They have large spins and, therefore, important energy barriers and small tunneling splittings. In other systems with small non-compensated spins (eventually zero in antiferromagnetic molecules), energy barriers are much smaller leading to much larger tunnel splitting. In this case, the mesoscopic physics is not limited to the spin bath, the phonons bath is also very relevant.

**Acknowledgments.**

We are very grateful to P.C.E. Stamp, N. Prokof'ev and W. Wernsdorfer for numerous helpful and motivating discussions. We thank L. Thomas, F. Lionti, I. Chiorescu, R. Giraud, A. Sulpice, C. Paulsen, A. Caneschi and D. Gatteschi for nice collaborations. We thank C. Paulsen and W. Wernsdorfer for providing us with originals of experimental figures. We are indebted to Ch. Barbara, A. Mishchenko and

S. Burmistrov for their assistance in the preparation of this manuscript. One from us (IT) is very grateful to J. Fourier University of Grenoble (France) for invitation and financial support during a stay in Grenoble. IT is also grateful to Laboratoire de Magnétisme Louis Néel (CNRS, Grenoble, France) and to Max-Plank Institute fur Festkorperforschung (CNRS, Grenoble, France) for its hospitality (and, personally, to I. Vagner and P. Wyder for everything what is absolutely impossible to list here due to the lack of space). This work was also supported by INTAS-97-12124 (European Community) and, partially, by RFBR-97-02-16548 (Russia).